\documentclass[letterpaper]{article} 
\usepackage{aaai25}  
\usepackage{times}  
\usepackage{helvet}  
\usepackage{courier}  
\usepackage[hyphens]{url}  
\usepackage{graphicx} 
\usepackage{natbib}  
\usepackage{caption} 
\usepackage{algorithm}
\usepackage[T1]{fontenc}    
\usepackage{url}            
\usepackage{booktabs}       
\usepackage{amsfonts}       
\usepackage{nicefrac}       
\usepackage{microtype}      
\usepackage{xcolor}         
\usepackage{tikz}
\usetikzlibrary{positioning}
\usepackage{float}
\usepackage{caption}
\usepackage{subcaption}
\usetikzlibrary {arrows.meta,automata,positioning}
\usepackage{algorithm}
\usepackage{algpseudocode}
\usepackage{amsmath}
\usepackage{amsfonts}
\usepackage{amssymb}
\usepackage{booktabs}
\usepackage{tabularx}
\usepackage{adjustbox}
\usepackage{caption}
\usepackage{array}
\usepackage{xspace}
\usepackage{comment}

\makeatletter
\newcommand{\multiline}[1]{%
  \begin{tabularx}{\dimexpr\linewidth-\ALG@thistlm}[t]{@{}X@{}}
    #1
  \end{tabularx}
}
\DeclareRobustCommand\onedot{\futurelet\@let@token\@onedot}
\def\@onedot{\ifx\@let@token.\else.\null\fi\xspace}

\definecolor{darkgreen}{rgb}{0.0, 0.5, 0.0} 
\definecolor{darkgreen}{rgb}{0.0, 0.5, 0.0}
\DeclareMathAlphabet\mathcal{OMS}{cmsy}{b}{n}
 
\algnewcommand{\LineComment}[1]{\State \(\triangleright\) #1}
\makeatother

\usepackage{newfloat}
\usepackage{listings}
\DeclareCaptionStyle{ruled}{labelfont=normalfont,labelsep=colon,strut=off} 
\floatstyle{ruled}
\newfloat{listing}{tb}{lst}{}
\floatname{listing}{Listing}
\title{Towards Efficient Collaboration via Graph Modeling in Reinforcement Learning}
\author{
    Wenzhe Fan\textsuperscript{\rm 1} $^\dagger$,
    Zishun Yu\textsuperscript{\rm 1},
    Chengdong Ma\textsuperscript{\rm 2},
    Changye Li\textsuperscript{\rm 3},
    Yaodong Yang\textsuperscript{\rm 2},
    Xinhua Zhang\textsuperscript{\rm 1}
}
\affiliations{
    University of Illinois Chicago \textsuperscript{\rm 1}\\
    Institute for Artificial Intelligence, Peking University\textsuperscript{\rm 2}\\ 
    Yuanpei College, Peking University\textsuperscript{\rm 3} \\
    wfan23@uic.edu,
    zyu32@uic.edu, 
    chengdong.ma@stu.pku.edu.cn, 
    antoine031106@gmail.com, \\
    yaodong.yang@pku.edu.cn,
    zhangx@uic.edu
}

\usepackage{bibentry}
\begin{document}

\maketitle
\renewcommand{\thefootnote}{\fnsymbol{footnote}}  
\footnotetext[2]{Correspondence to: Wenzhe Fan}

\begin{abstract}
In multi-agent reinforcement learning, a commonly considered paradigm is centralized training with decentralized execution. 
However, in this framework, decentralized execution restricts the development of coordinated policies due to the local observation limitation.
In this paper, we consider the cooperation among \textit{neighboring} agents during execution and formulate their interactions as a graph. 
Thus, we introduce a novel encoder-decoder architecture named Factor-based Multi-Agent Transformer ($f$-MAT) that utilizes a transformer to enable communication between neighboring agents during both training and execution. 
By dividing agents into different overlapping groups and representing each group with a \textit{factor}, $f$-MAT achieves efficient message passing and \textit{parallel} action generation through factor-based attention layers. 
Empirical results in networked systems such as traffic scheduling and power control demonstrate that $f$-MAT achieves superior performance compared to strong baselines, thereby paving the way for handling complex collaborative problems.
\end{abstract}

%
\section{Introduction}
The intricate nature of collaboration and the demand for time efficiency render multi-agent reinforcement learning (MARL) a challenging problem.
First, the joint action space grows exponentially with the number of agents, 
resulting in a complex scenario for making cooperative decisions. 
Second, it requires effective information exchange throughout the system to help agents learn the state of the environment and other agents.
For example, traffic light control at multiple intersections needs a coordination mechanism that allows each signal to act based on traffic conditions not only at its own intersection but also at nearby neighbors, even at distant signals.
Therefore, developing an efficient collaboration approach is crucial for decision-making in multi-agent systems.

A commonly considered paradigm in MARL is centralized training with decentralized execution (CTDE) \citep{sunehag2017value, rashid2020monotonic, son2019qtran, foerster2018counterfactual, lowe2017multi, yu2022the}, 
where each agent acts independently according to its own observation. 

However,
these methods only focus on stabilizing training with advanced value estimation and do not finely model the relationship among agents during execution.
As a result, some of them may fail in the simplest cooperative tasks \citep{kuba2022trust}.
Our method retains the actor-critic architecture in the CTDE framework
but extends independent action to neighborhood-based action, 
capturing the interactions between agents during execution.

There are literatures \citep{boehmer2020deep, li2020deep} modeling the relations among the agents in multi-agent systems with graph, in which each agent is represented by a node and the agent interaction is represented by an edge. 
To facilitate communication between agents, \citet{foerster2016learning} and \citet{ sukhbaatar2016learning} utilize the averaged encoded hidden states of other agents,
while \citet{zhang2018fully} and \citet{zhang2021decentralized} seek consensus to achieve the optimal common reward. 
\citet{das2019tarmac} and \citet{jiang2018learning} implemented attention mechanisms to determine the optimal time and target for communication. 
Further research employs graph neural networks (GNNs) and graph attention networks (GANs) to enhance agent interactions \citep{jiang2018graph, hoshen2017vain,das2019tarmac,singh2019individualized,niu2021multi,kim2019learning}.
Additionally, MARL methods capitalize on networked topologies \citep{chu2020multi,zhang2018fully,gupta2020networked,guestrin2002coordinated,zhang2007conditional}.
Unfortunately, many of these methods suffer from the time complexity of $O(n^2)$ for $n$ number of agents \citep{hao2023gat}. 
Moreover, agent-level communication limits the efficiency of learning cooperative policies.

To address these challenges, we propose Factor-based Multi-Agent Transformer ($f$-MAT), 
which enables efficient collaboration during \textit{both} training \textit{and} execution through all agents via graph modeling within the CTDE framework. 
First, to enable flexible communication, we propose a hypernode called $\textit{factor}$.
By modeling the collaboration structure as a graph, 
we organize agents into different groups and represent each of them as a factor. 
Serving as the intermediary, 
each factor can include multiple agents and each agent can belong to multiple factors.
Therefore, we allow an agent's observation and action to propagate and to influence other agents, while also facilitating communication at the group level.

Second, to capture the interactions between neighboring agents,
we utilize the transformer model, 
which shifts the search in the joint action space from a multiplicative to an additive size \citep{wen2022multi}, effectively addressing the problem of exponential growth of the action space. 
Furthermore, the mask mechanism ensures that messages are only passed between factors and their constituent agents.

Third, to address the $O(n^2)$ time complexity in GAN-based methods, 
we propose a factor-based attention that reduces the complexity to $O(m \cdot S_f \cdot L)$,
where $m$ is the number of factors,
$S_f$ is the maximum size of the factors (number of agents in it),
and $L$ is the number of layers.
Generally, this leads to significant savings as our experiments show.


Finally, to further reduce the computation cost, 
we propose a \textbf{parallel decoding} in transformer inference,
which significantly improves the efficiency upon the conventional autoregressive decoding.
This is particularly useful for MARL because the agents do not generally employ an order.

We evaluate the performance and efficiency of $f$-MAT in grid alignment, traffic scheduling, and power control.
Empirical results demonstrate that $f$-MAT fulfills the efficient collaboration compared to other baselines,
paving the way for efficient collaboration in multi-agent systems.


\section{Related Work}
Cooperative MARL presents a challenging problem as it is difficult for each agent to deduce its individual contribution to the global reward while cooperating with other agents.

A substantial amount of research has focused on the CTDE framework.
VDN \citep{sunehag2017value} directly factorizes the joint action-value function into the summation of independent Q-value functions. 
QMIX \citep{rashid2020monotonic} enforces the summation of action-value functions to a monotonic function. 
QTRAN \citep{son2019qtran} generalizes factorization by learning a state-value function, 
dispensing with the additivity and monotonicity assumptions.
COMA \citep{foerster2018counterfactual} introduces a counterfactual baseline in the advantage function and effectively isolates each agent's contribution.
MADDPG \citep{lowe2017multi} has access not only to the actions and observations of its corresponding agent but also to all other agents in the environment.
MAPPO \citep{yu2022the} is the first to apply PPO \citep{schulman2017proximal} to the multi-agent setting with parameter sharing.
These methods focus on learning a centralized critic, 
downplaying the interactions among agents, especially during execution. 
Recently, MAT \citep{wen2022multi} approaches MARL in a fully centralized fashion.
However, in practice, centralized execution is not feasible or is overly expensive in computation and communication.

To better model the relationship between agents, 
graphs have been commonly leveraged. 
DCG \citep{boehmer2020deep} considers a pre-specified coordination graph to enable message passing between agents and their neighbors. 
DICG \citep{li2020deep} improves this approach by inferring the dynamic coordination graph structure which is subsequently used by a GNN.
HAMA \citep{ryu2020multi} and MAGIC \citep{niu2021multi} utilize GANs to deal with communication between agents.
Although these works distill agent-agent interactions as edges in graph and take the direct neighbors into account, 
some of them still suffer from the time complexity of  \(O(n^2)\) problem for $n$ agents, 
especially in a dense graph or attention-based structure.
Nonetheless, the graph modeling approach greatly inspires us to leverage it for capturing more effective relationships among agents.

Further extensions allow agents to exchange messages during execution.
DIAL \citep{foerster2016learning} enables discrete communication via the limited-bandwidth channel. 
CommNet \citep{sukhbaatar2016learning} extends to a continuous communication channel.
TarMAC \citep{das2019tarmac} achieves targeted communication with a signature-based soft attention mechanism.
ATOC \citep{jiang2018learning} employs an attention mechanism
to decide whether an agent should communicate in its observable field.
NeurComm \citep{gupta2020networked} proposes a neural
communication protocol for networked system control.
However, these methods focus primarily on communication between individual agents and overlook communications at the group level, limiting their effectiveness in managing large-scale distributed systems. 
Yet, using attention mechanism to control when and with whom to communicate inspired us to employ transformers to ensure that information flows only to relevant agents.

In this paper, we begin with the CTDE framework and explore along the graph modeling perspective, 
aiming to find an efficient message-passing mechanism among agents during execution using the transformer model.

\section{Preliminaries}
\begin{figure*}[htb]
    \centering
    \begin{subfigure}[b]{0.15\textwidth}
        \centering
        \adjustbox{valign=c}{\includegraphics[width=0.6\linewidth]{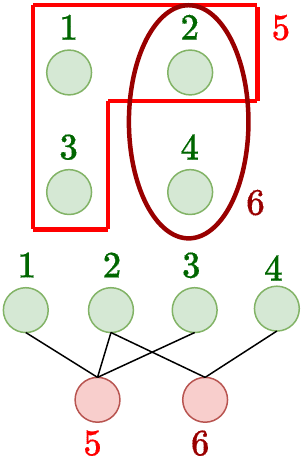}}
        \caption{A factor graph}
        \label{fig:factor_graph}
    \end{subfigure}
    \hfill 
    \begin{subfigure}[b]{0.42\textwidth}
        \centering
        \adjustbox{valign=c}{\includegraphics[width=0.6\linewidth]{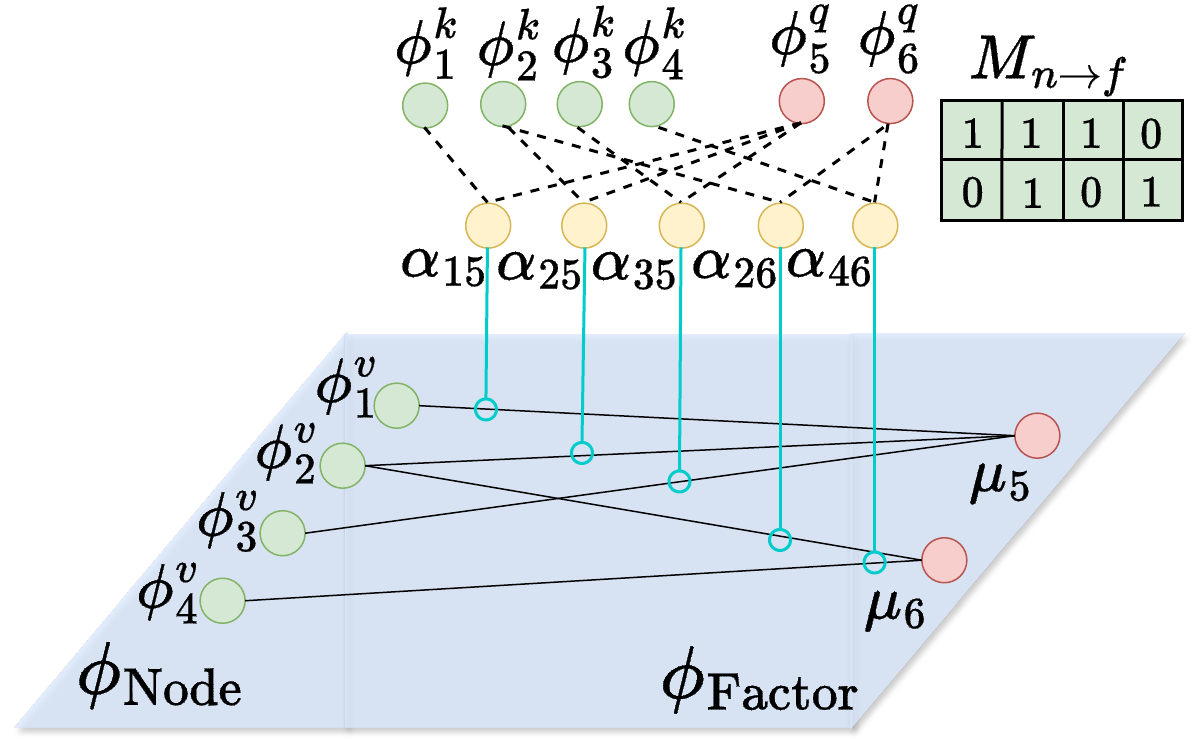}}
        \caption{Attention details when updating factors.}
        \label{fig:factor_based_attention}
    \end{subfigure}
    \hfill 
    \begin{subfigure}[b]{0.42\textwidth}
        \centering
        \adjustbox{valign=c}{\includegraphics[width=0.6\linewidth]{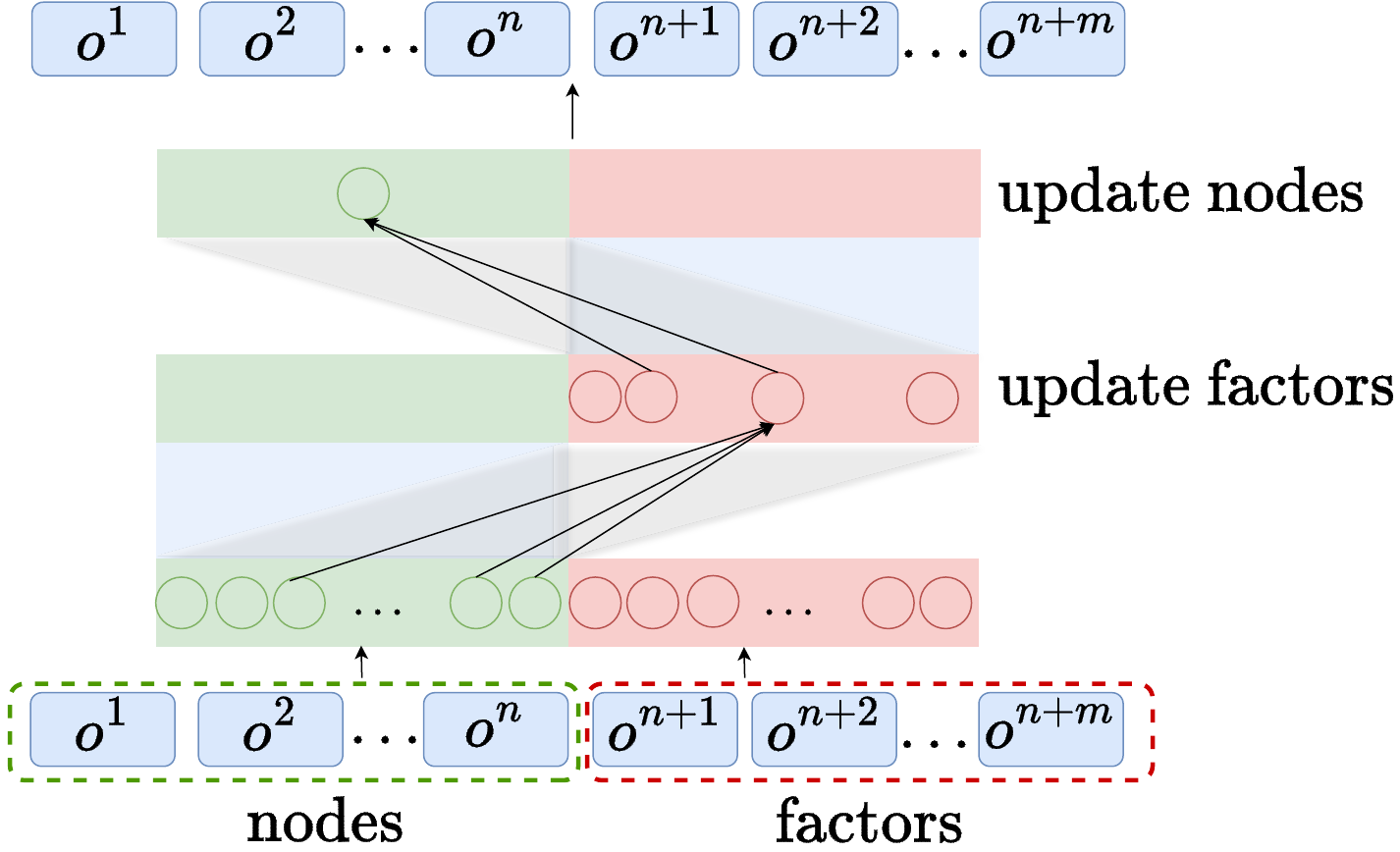}}
        \caption{Update scheme on factor-based MHA.}
        \label{fig:factor_update_structure}
    \end{subfigure}
    \caption{Factor-based attention layer. \textbf{\textcolor{darkgreen}{Green}} represents nodes, \textbf{\textcolor{red}{Red}} represents factors. (a) Factor graph: divide all nodes (1, 2, 3, 4) into two overlapping groups (1, 2, 3) and (2, 4); define two hypernodes, factor 5 and factor 6 to represent each group; transform a general graph to a bipartite graph.
    (b) Attention details when updating factors: to update factor observation $o_5, o_6$, we set $o_5, o_6$ as query, $o_1 \ldots o_4$ as key and value. Query $o^q_5$ only take attention to related agents' observations $o_1, o_2, o_3$. Similar operation to factor observation $o_6$. $\tilde{o}_5$ and $\tilde{o}_6$ are updated factors. 
    (c) Update scheme on factor-based MHA: It is a two-way message passing, which first updates factors and keep nodes unchanged, then update nodes and keep factors unchanged. }
    \label{fig:factor_attn}
\end{figure*}

We follow \citet{littman1994markov} to model cooperative MARL as a Markov game 
$\langle \mathcal{N}, \mathcal{O}, \mathcal{A}, \mathit{P}, \mathit{R}, \gamma \rangle$. 
$\mathcal{N} = \{ 1,2,3,..,n \}$ is the set of agents. 
$\mathcal{O}=\prod_{i=1}^n \mathcal{O}^i$ is the product of local observation spaces of the agents, namely, the agent observation space.
$\mathcal{A}=\prod_{i=1}^n \mathcal{A}^i$ is the product of the agents' action spaces, 
i.e., the joint action space. 
$\mathit{P}: \mathcal{O} \times \mathcal{A} \times \mathcal{O} \rightarrow $  $[0, 1] $ is the transition probability function. 
$\mathit{R}: \mathcal{O} \times \mathcal{A} \to \mathbb{R}$ is the joint reward function. $\gamma \in [0,1)$ is the discount factor.
Let $i_1, \ldots, i_n$ be a random permutation of $1, \ldots, n$,
and we abbreviate $i_{t:s} := \{i_t, i_{t+1}, \ldots, i_{s} \}$ if $t \le s$, and $\emptyset$ otherwise.

\subsection{Multi-Agent Transformer}
Multi-Agent Transformer \citep[MAT,][]{wen2022multi} casts cooperative MARL as a sequential modeling problem wherein one maps a sequence of observations to a sequence of actions, through the multi-agent advantage decomposition theorem \citep{kuba2021settling}.
This decomposition reveals the insight that the joint advantage $A^{i_{1:n}}_\pi$ can be decomposed into the sum of individual ones, allowing one to reduce the search space of multiplicative size $\prod^n_{i=1}|\mathcal{A}_i|$ to additive size $\sum^n_{i=1}|\mathcal{A}_i|$. 
In addition, the definition of individual advantage function $A_\pi^{i_m}(o, a^{i_{1:m-1}}, a^{i_m})$ naturally reveals a causal sequential structure, where $a^{i_m}$ depends on the set of preceding actions $a^{i_{1:m-1}}$ (and the joint observation). 
MAT leverages this decomposed causal structure, using encoder-decoder transformers with causal masked self-attention, by the following encoder and decoder training:
\begin{align*}
   L_{\text{enc}}(\phi) &= \frac{1}{Tn} \sum_{i=1}^{n} \sum_{t=0}^{T-1} \left[ \mathit{R}(\mathbf{o}_t,\mathbf{a}_t) + \gamma \mathit{V}_{\bar{\phi}} ({\hat{\mathbf{o}}}_{t+1}^{i}) - \mathit{V}_{\phi} ({\hat{\mathbf{o}}}_{t}^{i}) \right]   
     \\
     L_{\text{dec}}(\theta) &= 
     \frac{-1}{Tn} \!\! \sum_{m=1}^{n} \! \sum_{t=0}^{T-1} 
     \!\! \min \! \left( r_t^{i_m}(\theta) \hat{A}_t, \text{clip}(r_t^{i_m}(\theta), 1 \! \pm \! \epsilon) \hat{A}_t \right)  \displaybreak \\
   r_t^{i_m} (\theta) &= 
   \pi^{i_m}_{\theta}(a^{i_m}_t | \hat{\mathbf{o}}_t^{i_{1:n}}, \hat{\mathbf{a}}_t^{i_{1:m-1}}) \ / \  
   \pi^{i_m}_{\theta_\text{old}}(a^{i_m}_t | \hat{\mathbf{o}}_t^{i_{1:n}}, \hat{\mathbf{a}}_t^{i_{1:m-1}}) 
\end{align*}
The policy of agent $i_m$ is $\pi^{i_m}_{\theta}(a^{i_m}_t | \hat{o}_t^{i_{1:n}}, \hat{a}_t^{i_{1:m-1}})$,
requiring observation representations of all agents $\hat{o}_t^{i_{1:n}}$ and all preceding agents' actions $\hat{a}_t^{i_{1:m-1}}$.
So the execution is centralized. 
In practice, decision-making may not require complete information from the system.
Often, only observations from nearby or related neighbors are relevant.
Pulling information from all agents can introduce redundant details, 
wasting computation and communication. 
That said, MAT inspires a transformer-based structure in our message-passing mechanism.

\section{Factor-based Multi-Agent Transformer}
The goal of $f$-MAT is to address the challenge of multi-agent collaboration in execution for centralized training decentralized execution (CTDE) algorithms, aiming to generate more cooperative policies. 
In this paper, we focus on exploring a message-passing mechanism from a graph modeling perspective, seeking to enhance cooperation through a more efficient and expansive communication approach.

\subsection{Factor Representation of Coordination Graph}
\label{sec:factor_graph}

To enable the broader message passing, we divide agents into different groups and represent each group with a virtual hypernode named \textit{factor}, thereby using factors as the intermediary to fulfill the group-level communication.  
Pooling several agents into one group instills the prior that they tend to influence each other.
For example, cooperation is particularly necessary for them to achieve optimal actions, 
or an agent's optimal policy should draw on a factor-mate's inputs (not necessarily the raw observations),
or agents in this factor collectively define some situations that impact other agents.
We will pass low-cost messages to agents within the same factor to share information.
Therefore, a larger factor promotes collaboration between more agents.
Since an agent can belong to multiple factors,
larger factors also allow information such as actions and observations to be propagated more efficiently to other agents.

We use the toy example in Fig.~\ref{fig:factor_graph} to illustrate our idea,
and it can be easily extended to general networks. 
Factors can be defined flexibly, 
accounting for multiple inductive biases and practical constraints.
For example, it can be any group of agents such as (1, 2), (2, 3), (3, 4) or (1, 2, 3, 4). 
Here, we divide the graph into two groups: (1, 2, 3) and (2, 4), represented by factors 5 and 6, respectively.

Following the standard practice in the graphical model literature \citep{Bishop06}, 
we characterize the agent-factor membership with a bipartite graph $\mathcal{G} = \langle \mathcal{N}, \mathcal{F} \rangle$, 
where $\mathcal{F} = \{f_j : j = 1, \ldots, m\}$ is a set of factors. 
An undirected edge is placed between a node $i \in \mathcal{N}$ and a factor $f \in \mathcal{F}$ if and only if $i$ is a member of $f$, denoted as $i \in f$.
Denote the set of edges as $\mathcal{E}$.
Fig.~\ref{fig:factor_graph} shows the resulting factor graph.

The factor representation simplifies the complex collaboration among agents on a general graph into a group-level message passing framework represented by a bipartite graph, which can be easily utilized by the transformer architecture. 
In the next section, we will overcome the $O(n^2)$ complexity of transformer and explore an efficient message passing mechanism across the entire graph via factors. 
\subsection{Factor-based Attention Layer}
\label{sec:f-MHA}
\begin{figure*}[!htb]
    \centering
        \includegraphics[width= 0.6\linewidth]{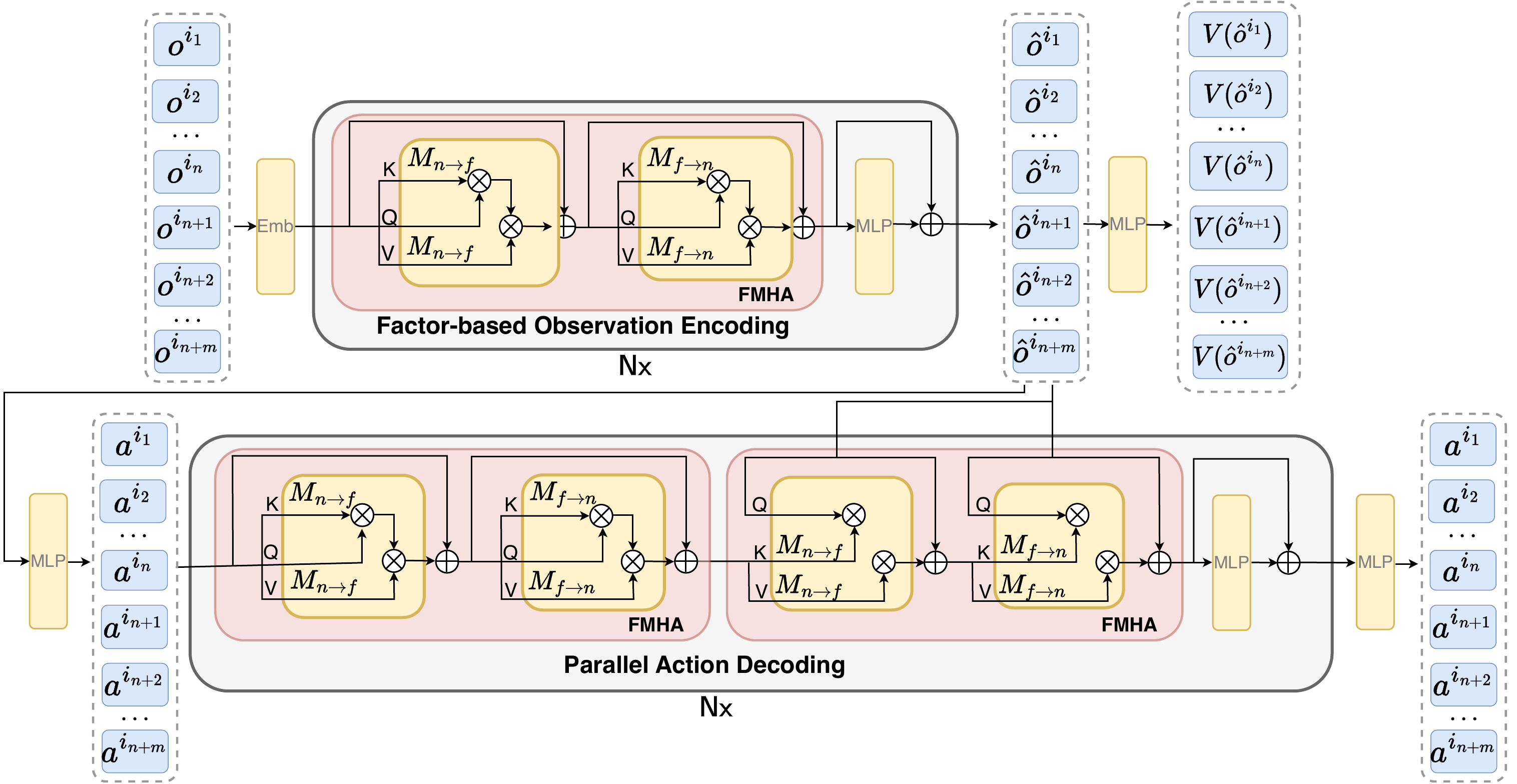}
        \caption{Architecture of $f$-MAT. 
        At each time step, the encoder takes the observation of nodes and factors as the input and outputs the factor-based observation representation. The factor observation is initialized by the average of related agent's observations. In decoder, we initialized the actions by the learned observation representation and generate actions in parallel. All attention layers utilized are factor-based attention layers.
        The pseudo code of $f$-MAT can be found in Appendix \ref{sec:pseudo}.}
        \label{fig:f_MAT}
\end{figure*}
A crucial property of $f$-MAT is passing messages between agents and factors.
We propose factor-based attention layer to fulfill the efficient message passing by combining factor and multi-head attention layers (MHA) in transformer through proper masking.

Assume we have $n$ nodes, $m$ factors, 
the input sequence matrix is $O \in \mathbb{R}^{(n+m) \times D}$, 
and $D$ is the embedding dimension. 
As illustrated in Fig. \ref{fig:factor_update_structure}, 
$O[\mathcal{N},:]$ with $\mathcal{N} \in [1,\dots,n]$ is the raw observation of nodes, 
and $O[\mathcal{F},:]$ with $\mathcal{F} \in [n+1, \dots, m]$ is the observation of factors initialized by averaging the observations of constituent nodes. 
We calculate factor embeddings using nodes, 
followed by computing node embeddings using factors. 
During message passing, 
masks are employed to ensure each factor only pays attention to its connected nodes, 
and each node only pays attention to its connected factors. 
To maintain the fixed length observation $n+m$, 
we keep the nodes unchanged when updating the factors,
and the factors unchanged when updating the nodes.

For example,
in Fig. \ref{fig:factor_graph}, 
nodes are defined as $\mathcal{N}=\{1, \ldots, 4\}$, 
factors are defined as $\mathcal{F}=\{5, 6\}$,
and the input of the factor-based model is
$O[\mathcal{N} \cup \mathcal{F},:]=(o^1, \ldots, o^4, o^5, o^6)^\top$. 
In the attention layer that sends messages from nodes to factors, 
we first keep nodes ($o^1, \ldots, o^4)$ unchanged, then update factors by masking out all unrelated nodes. 
Assume the updated factors are denoted as $(\tilde{o}^{5}, \tilde{o}^{6})$, 
and the output of the attention layer with mask $M_{n \to f}$ (meaning node to factor) is $(o^1, \ldots, o^4, \tilde{o}^{5}, \tilde{o}^{6})$.
In the attention layer that sends messages from factors to nodes, we first keep factors $(\tilde{o}^{5}, \tilde{o}^{6})$ unchanged, then update nodes by masking out all unrelated factors. 
Assume the updated nodes are denoted as $(\tilde{o}^{1}, \ldots, \tilde{o}^{4})$, 
and the output of the attention layer with mask $M_{f \to n}$ is $(\tilde{o}^1, \ldots, \tilde{o}^4, \tilde{o}^5, \tilde{o}^{6})$.

Here, we defined two mask matrices of $(n+m) \times (n+m)$: 
$M_{n \to f}$ for message passing from nodes/agents to factors/groups,
and its \textit{transpose} as $M_{f \to n}$ for the opposite direction:
\begin{align}
    M_{n \to f}(i,j) &= 
    \begin{cases}
        1 & \text{if } i \in \mathcal{N} \text{ and } j \in \mathcal{F} \text{ and } i \in f_j \\
        0 & \text{else}
    \end{cases}.
\end{align}

Both matrices are sparse,
rendering optimization in implementation.
Fig. \ref{fig:factor_based_attention} shows the top-right corner of
$M_{n \to f}$ for the running example, i.e., rows $1$ to $n$ and columns $n+1$ to $n+m$;
the other elements are 0.
The queries $Q$ are no longer represented by $O \in \mathbb{R}^{(n+m) \times D}$. 
Instead, in factor-based attention, queries $Q$ are set to $O[\mathcal{F},:]$, 
keys $K$ to $O[\mathcal{N},:]$, and values $V$ to $O[\mathcal{N},:]$. 
Similar operations apply to $M_{f \to n} $.

Applying the factor-based attention mechanism to our running example,
the queries are $Q=\{5, 6\}$, with keys $K$ and values $V$ being $\{1,2,3,4\}$. 
When updating factors, each factor only attends to its constituent nodes. 
Thus, $o^5_\text{q}$ only attends to $\{o^1_\text{k}, o^2_\text{k}, o^3_\text{k}\}$, and $o^6_\text{q}$ only attends to $\{o^2_\text{k}, o^4_\text{k}\}$. 
Once factors $o^5$ and $ o^6$ are updated, 
we maintain the fixed-length observation $n+m$ without changing those for the nodes.

The update of nodes and factors is completed after passing through two layers of factor-based attention (node-to-factor and factor-to-node). We define it as factor-based multi-head attention ($f$-MHA),
and multiple $f$-MHA layers allow longer-distance propagation between agents through factors, so that all the necessary information is spread throughout the graph.  

The time and space complexity of the factor-based attention model with $L$ layers is $O(L \cdot |\mathcal{E}|)$,
which is bounded by $O(m \cdot S_f \cdot L)$.
Typically, $L \leq 3$. 
Depending on the factor topology,
it can be much more efficient than the traditional attention models,
which cost $O(n^2)$.
We will demonstrate the savings via the three experiments in Sec.~\ref{sec:experiments}.


\subsection{Encoder and Decoder Implementations}

The overall architecture of $f$-MAT can be found in Fig.~\ref{fig:f_MAT}.
Our implementation is based on MAT, where we replaced conventional MHA with $f$-MHA, leading to a few nuances in the encoder/decoder updates and inference.

\noindent {\bf Encoder} \quad In conventional encoder-decoder transformers as well as in MAT, attention is not masked in encoders, leading to centralized policy w.r.t. observations, i.e., requiring joint observation $\mathbf{o}$ as the input. 
In contrast, $f$-MAT applies factor-based masks to enable local message passing, 
so that a policy only needs to draw upon \textcolor{blue}{local observations} instead of global ones. 
Equation \eqref{eq:fmat_encoder} below formalizes the encoder objective of $f$-MAT,
using the Bellman error of the value function $V$ that is defined locally.
The pseudocode is given in Algorithm~\ref{alg:encoder}.
\begin{equation}
\label{eq:fmat_encoder}
   L_{\text{Encoder}}(\phi) = \frac{1}{Tn} \! \sum_{i=1}^{n} \! \sum_{t=0}^{T-1} 
   \!\! 
   \left[ \mathit{R}(\mathbf{o}_t,\mathbf{a}_t) \! + \! \gamma \mathit{V}_{\bar{\phi}} ({\hat{\mathbf{o}}}_{t+1}^{i})  
   \! - \! \mathit{V}_{\phi} ({\hat{\mathbf{o}}}_{t}^{i}) \right]
\end{equation}

\noindent{\bf Decoder} \quad Similarly, decoders are composed of our factor-based attention layers, and training is similar to conventional decoder training, except that our attention is local. 
\textbf{The major difference lies in inference}, the computational bottleneck of MAT due to the auto-regression. 
Furthermore, since there is generally no natural order among agents,
MAT manually introduces a random order and regenerates it every iteration.


In contrast,
we propose \textbf{parallel inference} for $f$-MAT which is by itself novel for transformers. 
The method is detailed in Algorithm~\ref{alg:decoder} in Appendix \ref{sec:pseudo},
where message passing alternates between node-to-factor and factor-to-node in a \textbf{synchronized} fashion.
This resembles Gibbs sampling,
and we sketch the connection in Appendix~\ref{sec:gibbs_app}.

We observe that a small number of $f$-MHA layers ($L \leq 3$) can already produce good performance in our experiments.
In addition, we consider using action distributions (directly maps the factored observations to individual actions through linear layers) as initialization to bypass the slow mixing issue of Gibbs, as it is known that good initialization improves efficiency of samplers~\citep{boland2018efficient}. 

For training, 
we update the decoder using PPO loss \citep{schulman2017proximal} with a factorized policy:
\begin{align}
\label{eq:decoder}
   L_{\text{Decoder}}(\theta) &\! =\! \frac{-1}{Tn} \! \sum_{i=0}^{n} \! \sum_{t=0}^{T-1} \! \min \! \left( r_t^i(\theta) \hat{A}_t, \text{clip}(r_t^i(\theta), 1 \pm \epsilon) \hat{A}_t \right)
   \\
    r_t^i (\theta) &\! = \! 
    \pi^i_{\theta}(a^i_t | \hat{\mathbf{o}}_t^{{\color{blue}\text{rf}(i)}}) \ / \     \pi^i_{\theta_{\text{old}}}(a^i_t | \hat{\mathbf{o}}_t^{{\color{blue}\text{rf}(i)}}).
\end{align}
The policy $\pi^i$ for agent $i$ draws on the observations from its {\color{blue} reception field} rf$(i)$.
It consists of all the agents that can be reached from $i$ with at most $2L$ hops on the factor graph.

The complete pseudocode for $f$-MAT's encoder and decoder can be found in Algorithm~\ref{alg:ours_overall} in Appendix~\ref{sec:pseudo}. 
$f$-MAT seeks a balance between centralized and decentralized execution, offering a novel solution to cooperative MARL. 
The key insight is the factor-based mechanism, which facilitates efficient and extensive message passing among agents from graph modeling perspective.
Additionally, $f$-MAT enables parallel action generation within the transformer model, further reducing computational time and making it better suited for environments that require agents to take cooperative actions simultaneously.

\section{Experiments}
\label{sec:experiments}

We evaluate $f$ -MAT in three environments, 
each presenting different challenges.
The first environment is a strong cooperation scenario named grid alignment, 
where the agents prefer to align their actions with the neighbors of the same row or column.
The second environment is traffic light control with heterogeneous agents.
The third environment is power control, emphasizing local control where an agent's actions have a limited effect on those distant from it.
We choose MAT, a fully centralized method, as the performance upper bound.
Then, we primarily utilize MAPPO and MAT-dec\footnote{MAT-dec is a more decentralized variant of MAT, but it still has a centralized component at execution, namely the encoder.} under the CTDE framework as our baseline competitors.
We evaluate all methods in terms of performance and training efficiency.


\subsection{Grid Alignment}
\label{sec:grid_align}

\noindent \textbf{Environment} \quad Our first experiment is on a simplified domain of traffic flow \citep{zhang2007conditional}, 
called \textbf{GridSim}, where all agents coordinate their actions to maximize the global reward. 
The traffic flow is an $s \times s$ grid shown in Fig.~\ref{fig:env_gridsim} in Appendix~\ref{sec:app_env}. 
Each row and column includes a buffer to hold traffic units that arrive with a probability ($\text{Pr}=0.5$) at each timestep. 
At each grid intersection, an agent controls a gate,
which can be chosen from two actions:
keeping the gate aligned horizontally or vertically. 
Traffic can only flow through a column if \textbf{all} its gates are vertically aligned, 
and similarly through a row if all its gates are horizontally aligned. 
When this happens, all waiting traffic for that line
propagates instantly, and each unit of traffic contributes $+1$ to a global reward.
The optimal reward is the grid size $s$. 
So each agent should ideally choose the same actions as its directly preceding neighbor, especially if that neighbor is in the lane with the most waiting traffic in the buffer. Hence, this environment emphasizes the collaboration of direct neighbors.

\begin{figure}[t]
    \centering
    \begin{subfigure}[b]{0.23\textwidth}
        \centering
        \includegraphics[width=0.9\textwidth]{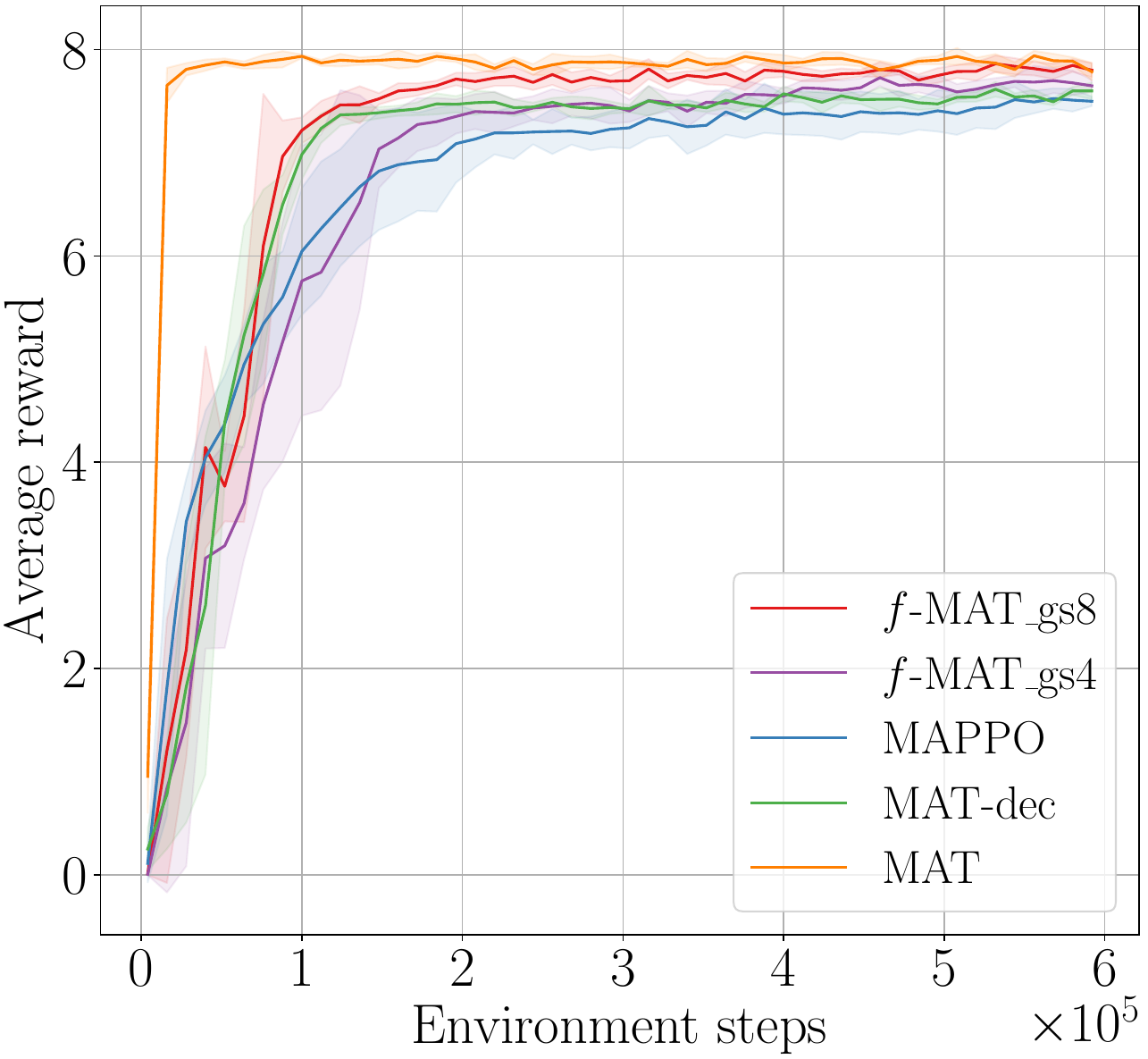}
        \caption{Performance on $8 \times 8$ grid}
        \label{fig:grid8_per}
    \end{subfigure}
    \hfill
    \begin{subfigure}[b]{0.23\textwidth}
        \centering
        \includegraphics[width=0.9\textwidth]{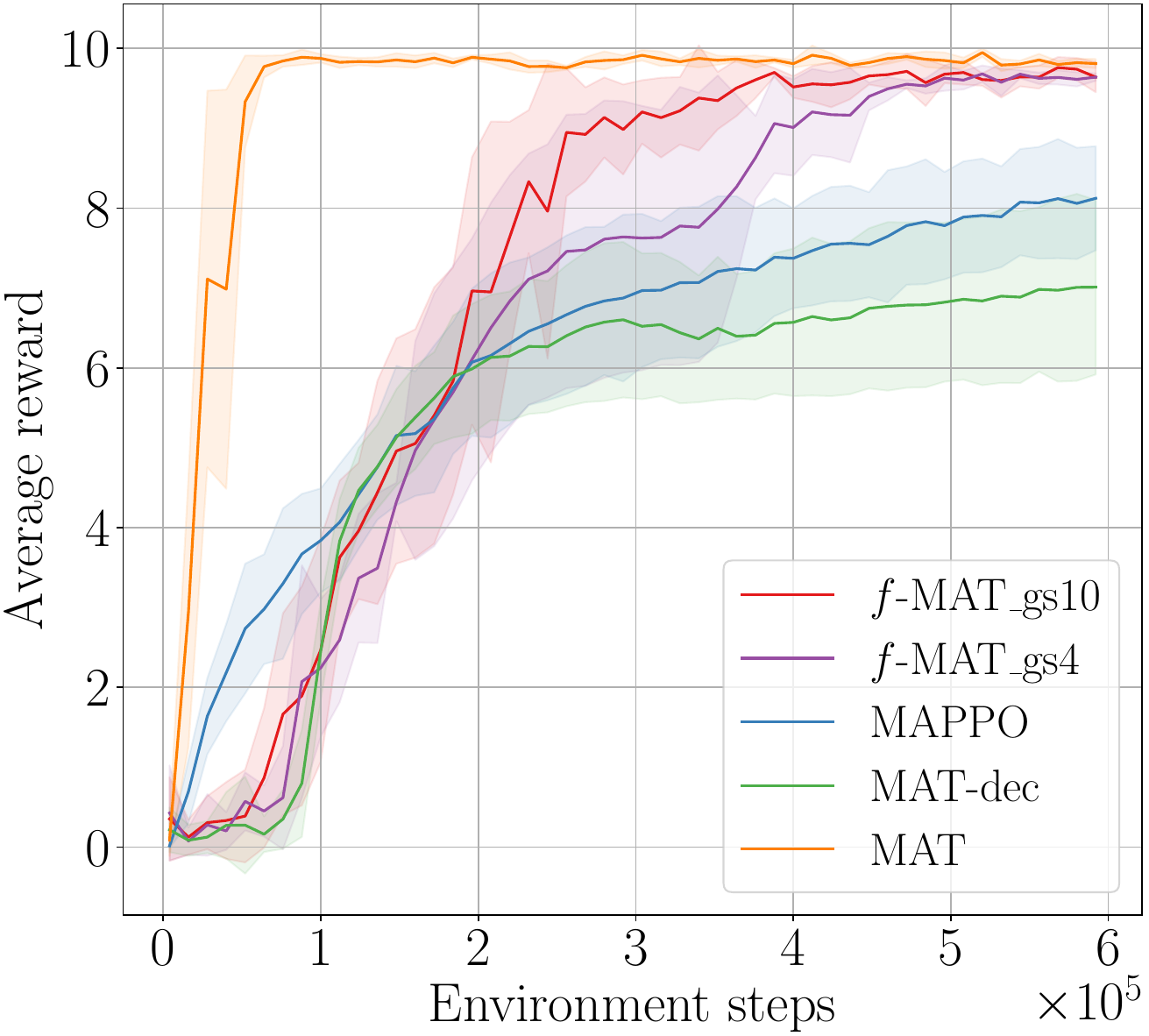}
        \caption{Performance on $10 \times 10$ grid}
        \label{fig:grid10_per}
    \end{subfigure}
    \vfill
    \begin{subfigure}[b]{0.23\textwidth}
        \centering
        \includegraphics[width=0.9\textwidth]{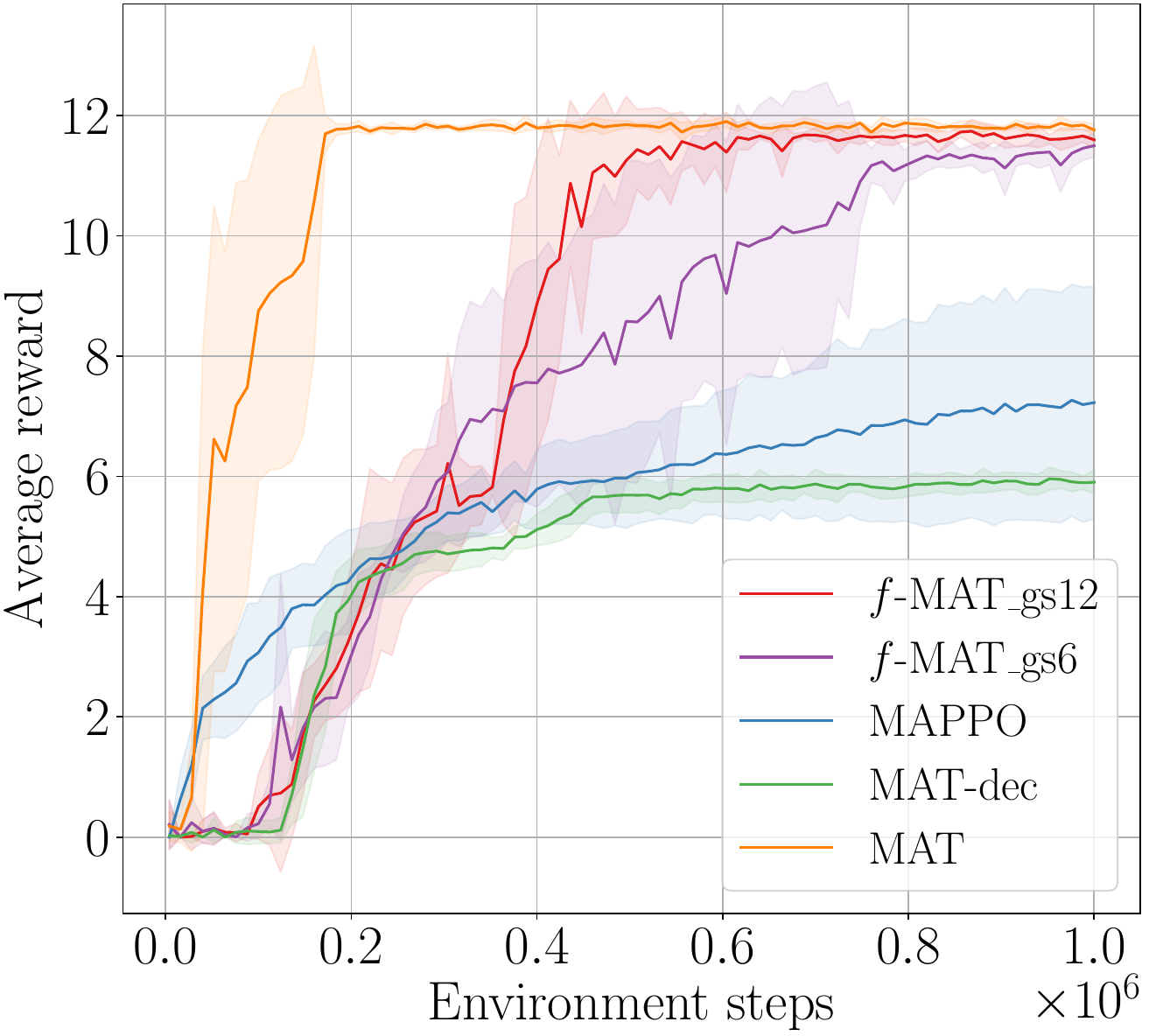}
        \caption{Performance on $12 \times 12$ grid}
        \label{fig:grid12_per}
    \end{subfigure}
    \hfill
    \begin{subfigure}[b]{0.23\textwidth}
        \centering
\includegraphics[width=0.9\textwidth]{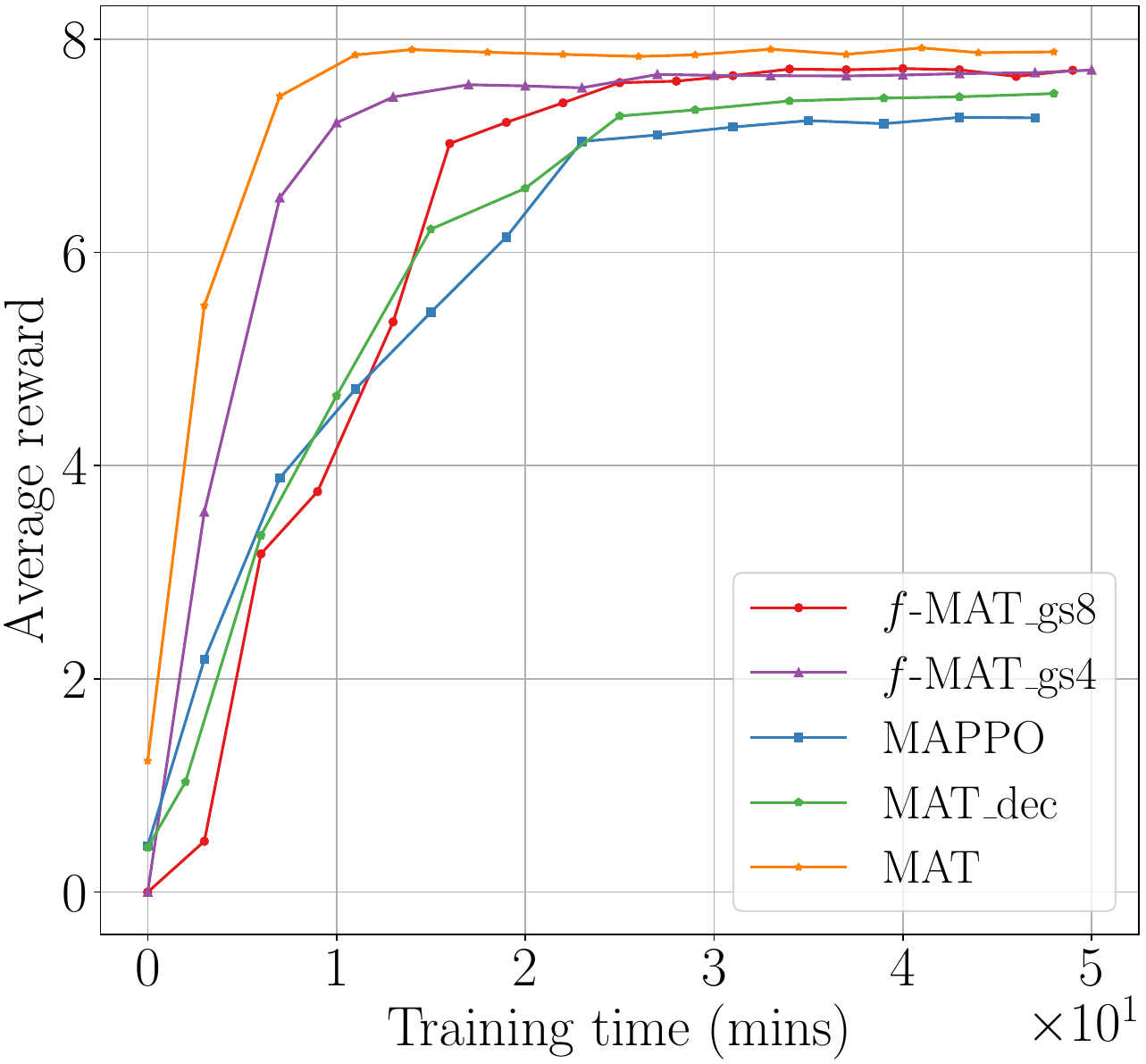}
        \caption{Efficiency on $8 \times 8$ grid}
        \label{fig:grid_efficieny}
    \end{subfigure}
    \caption{The performance results for \textbf{GridSim} with three different number of agents ($n=64, 100, 144$) and the training efficiency on grid $8 \times 8$. All performance results are presented as mean $\pm$ std. 
    `Efficiency' in subplot (d) refers to training efficiency, i.e., evaluation reward vs. training time. }
    \label{fig:gridSim_result}
\end{figure}

\noindent
\textbf{Results}  \quad  
It is natural to form groups/factors along each row and column.
Figure \ref{fig:gridSim_result} illustrates the resulting performance and training efficiency.
The `gs' in legends refers to the group size ($S_f$) - `gs4' means that each factor involves $S_f = 4$ agents located consecutively along each row or column.
This leads to $O(s (s - S_f) S_f L)$ complexity,
which is much lower than $O(n^2) = O(s^4)$.

MAT sets an upper bound in GridSim where an agent's action depends on the actions of preceding neighbors. Comprehensive knowledge of all preceding agents aids decision-making. 
In the $8 \times 8$ grid shown in Fig. \ref{fig:grid8_per}, while all CTDE methods achieve optimal results, 
$f$-MAT is the closest to MAT in performance and converges the fastest.
As we move towards a \(12 \times 12\) grid, 
the differences in performance among the four methods become more pronounced. 
\(f\)-MAT continues to approach the optimal performance, 
whereas MAT-dec and MAPPO fall short,
achieving 75\% and 65\% of $f$-MAT's performance on 
$10 \times 10$ (Fig. \ref{fig:grid10_per}) and $12 \times 12$ grids (Fig. \ref{fig:grid12_per}), 
respectively. 
Interestingly, Fig. \ref{fig:grid10_per} and \ref{fig:grid12_per} show that the group size significantly impacts the global reward. 
We will delve into the impact of group size in Sec.~\ref{sec:alb}.

To compare the training efficiency,
Fig.~\ref{fig:grid_efficieny} shows that $f$-MAT achieves optimal performance among all CTDE methods, just slightly slower than MAT.
$f$-MAT requires only 1/2 of the training time taken by MAPPO and MAT-dec to reach the similar performance.
This confirms that $f$-MAT is significantly more efficient in learning.
We note that this subplot is based on a single seed,
because it is difficult to plot the average over multiple seeds.
We hence present the results of two more seeds in Figure~\ref{fig:supp_grid8_time} in Appendix~\ref{sec:other_results_app}.
Similar results with additional seeds for traffic control and power control are in Fig.~\ref{fig:supp_monaco_time} and Fig.~\ref{fig:supp_power_time}, respectively.

\subsection{Traffic Light Control}
\label{sec:experiment_traffic}
\noindent \textbf{Environment} \quad 
The second environment adapted the Simulation of Urban Mobility \citep[SUMO,][]{ chen2020toward, ault2021reinforcement}, 
which is widely recognized in the transportation community.
We chose as our testbed an area with $n=28$ traffic lights in Monaco \citep{chu2020multi},
illustrated in Fig. \ref{fig:env_monaco} in Appendix~\ref{sec:app_env}.
We used the average queue length at intersections to measure the level of traffic congestion.
Traffic light control is a challenging application in MARL because each traffic light needs to observe a wider area to make decisions, 
and the actions of each agent impact a broader region beyond just adjacent agents.
Another challenge arises from the \textit{heterogeneity} of agents in terms of observation and action.
\begin{figure}[t]
    \centering
    \begin{subfigure}[b]{0.23\textwidth}
        \centering
\includegraphics[width=\linewidth]{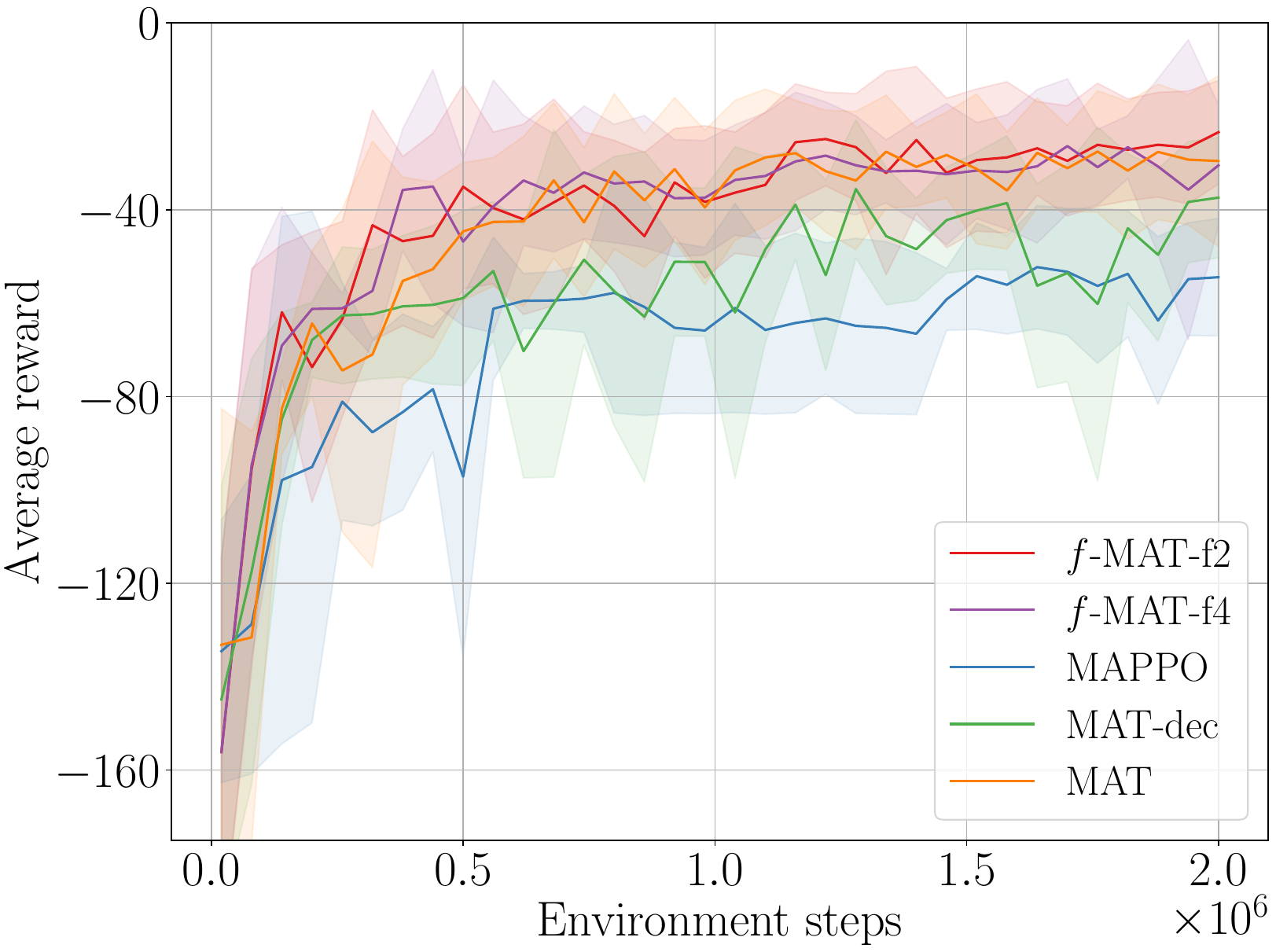} 
        \caption{Performance on Monaco}
        \label{fig:monaco_per}
    \end{subfigure}%
    \hfill
    \begin{subfigure}[b]{0.23\textwidth}
        \centering
        \includegraphics[width=\linewidth]{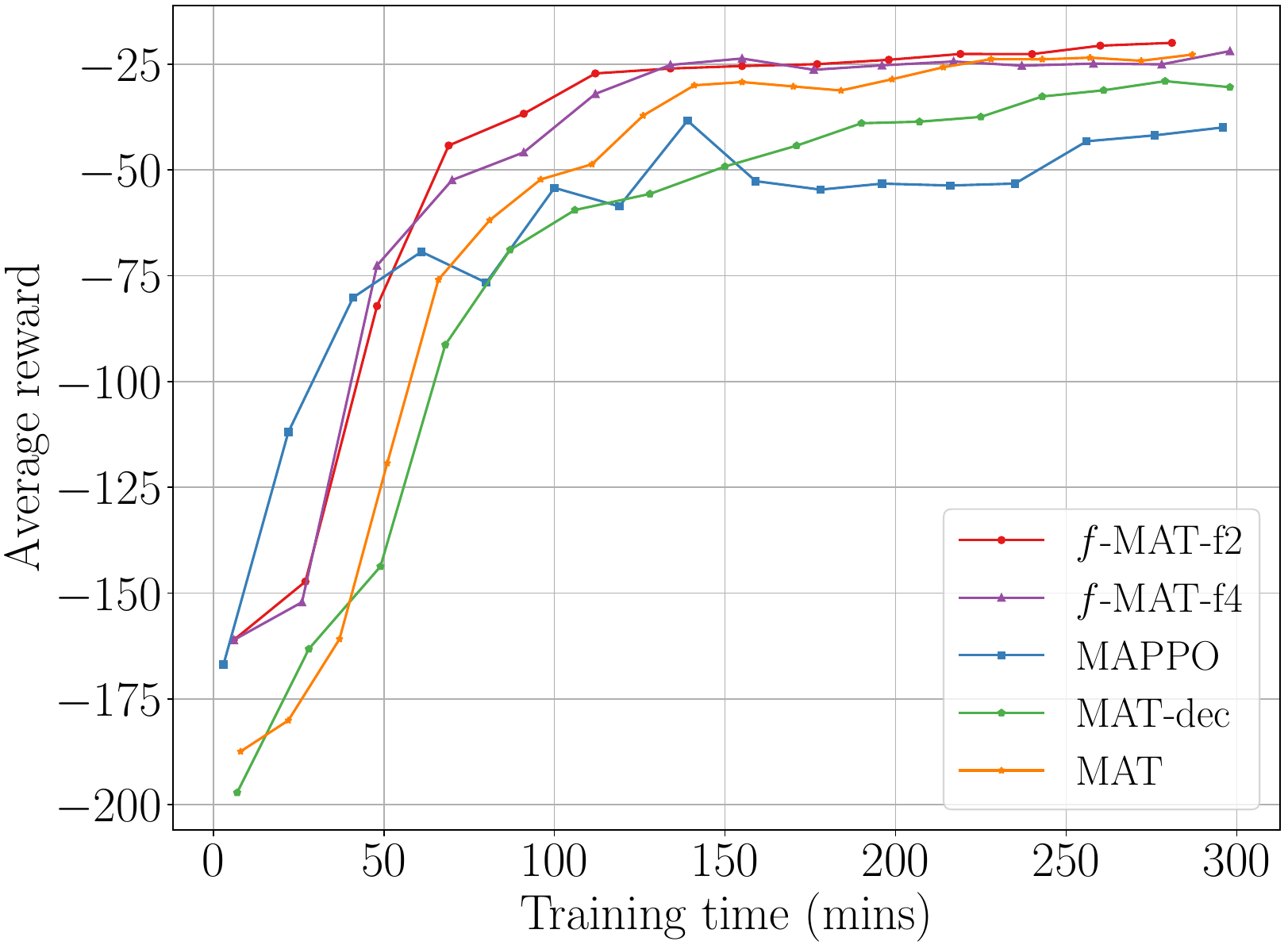} 
        \caption{Efficiency on Monaco}
        \label{fig:monaco_efficiency}
    \end{subfigure}
    \caption{The performance and training efficiency results for \textbf{traffic light control}, an area in Monaco with 28 traffic lights. $f$-MAT is compatible with \textit{heterogeneous} environments.}
    \label{fig:monaco_result}
\end{figure}

\noindent
\textbf{Results}  \quad  
It is less clear here how to form the groups based on the problem formulation or the definition of the reward.
We randomly divided the agents into two groups or four groups,
and then we added each agent's directly connected neighbors.
This leads to $f$-MAT-f2 with $S_f=20$ and $f$-MAT-f4 with $S_f=15$ approximately.
The complexity of $O(S_fL)$ is much lower than $O^2$.
We will discuss group selection later in Sec.~\ref{sec:alb}.

Fig.~\ref{fig:monaco_result} shows that under different numbers of factors, $f$ -MAT produces performance comparable to MAT and achieves a higher training efficiency than all other methods.

Furthermore, $f$-MAT maintains its effectiveness even in a heterogeneous environment. 
In contrast, MAPPO, being a shared parameter approach, is inherently susceptible to failure in an inhomogeneous setting. 
MAT, MAT-dec, and $f$-MAT incorporate an embedding layer that aligns the diverse observation and action dimensions, thereby enabling their adaptability and success in heterogeneous environments. 

Given the need for collaboration over a broader area and communication between agents in this environment, 
$f$-MAT outperforms MAT-dec in both performance and training efficiency. 
$f$-MAT learns faster than MAT, requiring 3/5 training time of MAPPO and MAT-dec to achieve comparable results. 
This highlights the need for cooperation during execution and further validates the advantage of $f$-MAT.

\subsection{Power Control}
\noindent \textbf{Environment} \quad The voltage control problem in distributed generators (DGs) can be viewed as a cooperative MARL problem. 
We have two microgrid systems \citep{chen2021powernet}: one with 6 distributed DGs (microgrid-6) 
and a larger-scale microgrid system with 20 DGs (microgrid-20), 
both shown in Fig. \ref{fig:env_power} in Appendix~\ref{sec:app_env}. 
Power control is an environment where communication among all agents is not necessary, as control infrastructures are typically dispersed across a large area.
It is a widely used environment for communication-based methods. 
We reference their results and compare them with our methods in the Table \ref{table:power} in Appendix~\ref{sec:other_results_app} .

\begin{figure}[t]
    \centering
    \begin{subfigure}[b]{0.23\textwidth}
        \centering
        \includegraphics[width=0.9\linewidth]{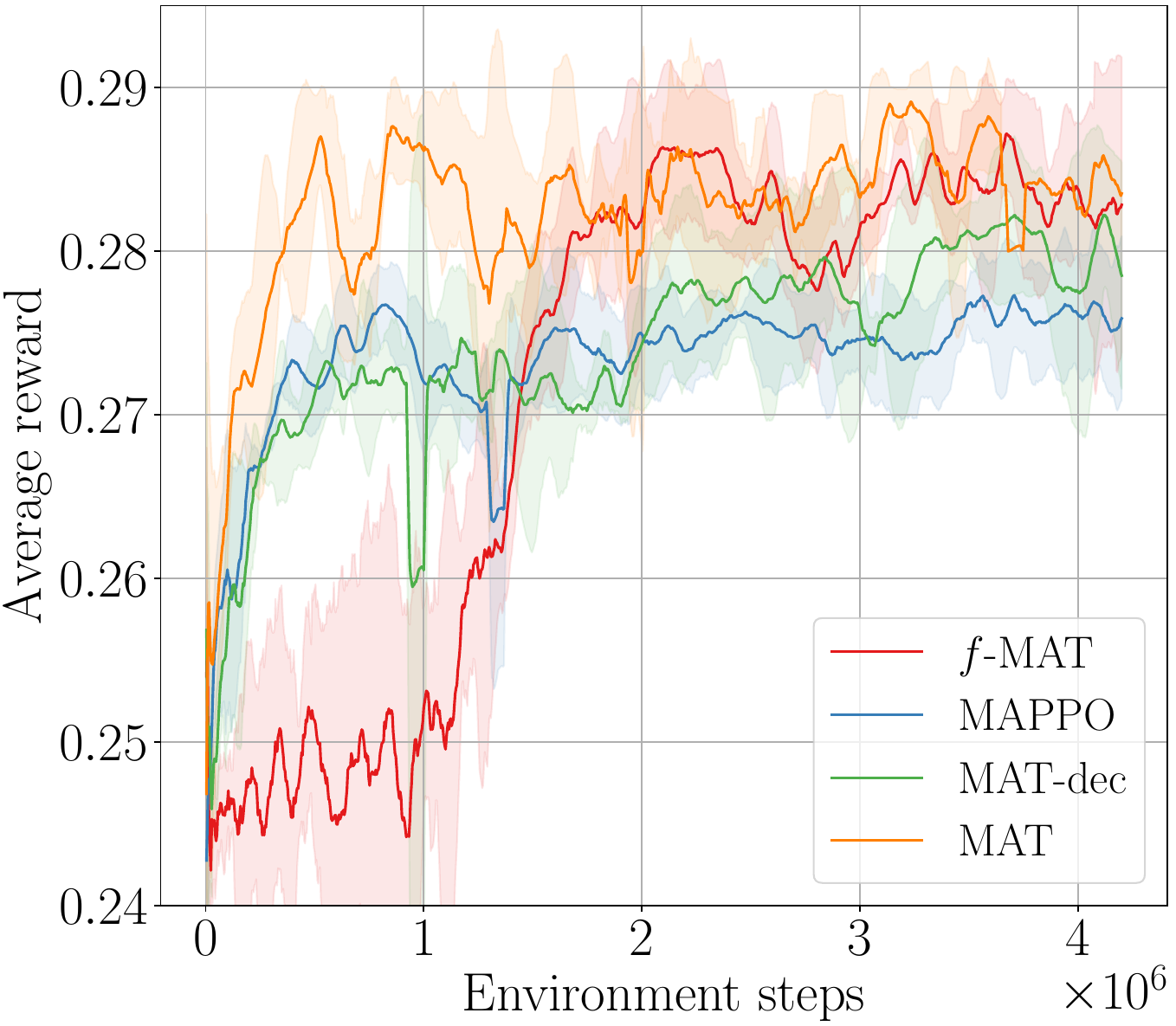} 
        \caption{Performance on microgrid-6}
        \label{fig:power6_per}
    \end{subfigure}%
    \hfill
    \begin{subfigure}[b]{0.23\textwidth}
        \centering
        \includegraphics[width=0.9\linewidth]{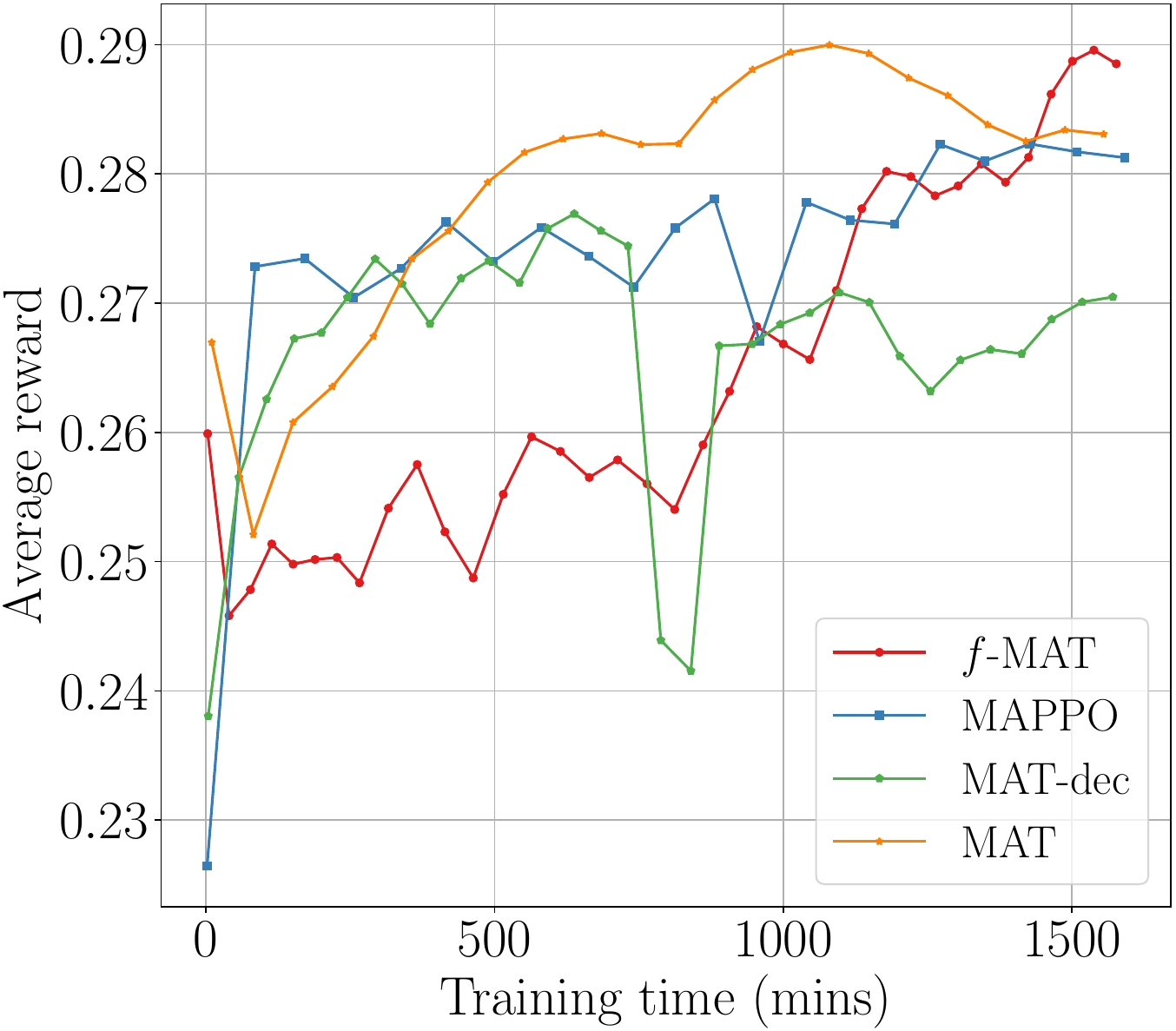} 
        \caption{Efficiency on microgrid-6}
        \label{fig:power6_efficiency}
    \end{subfigure}
    \vfill
    \begin{subfigure}[b]{0.23\textwidth}
        \centering
        \includegraphics[width=0.9\linewidth]{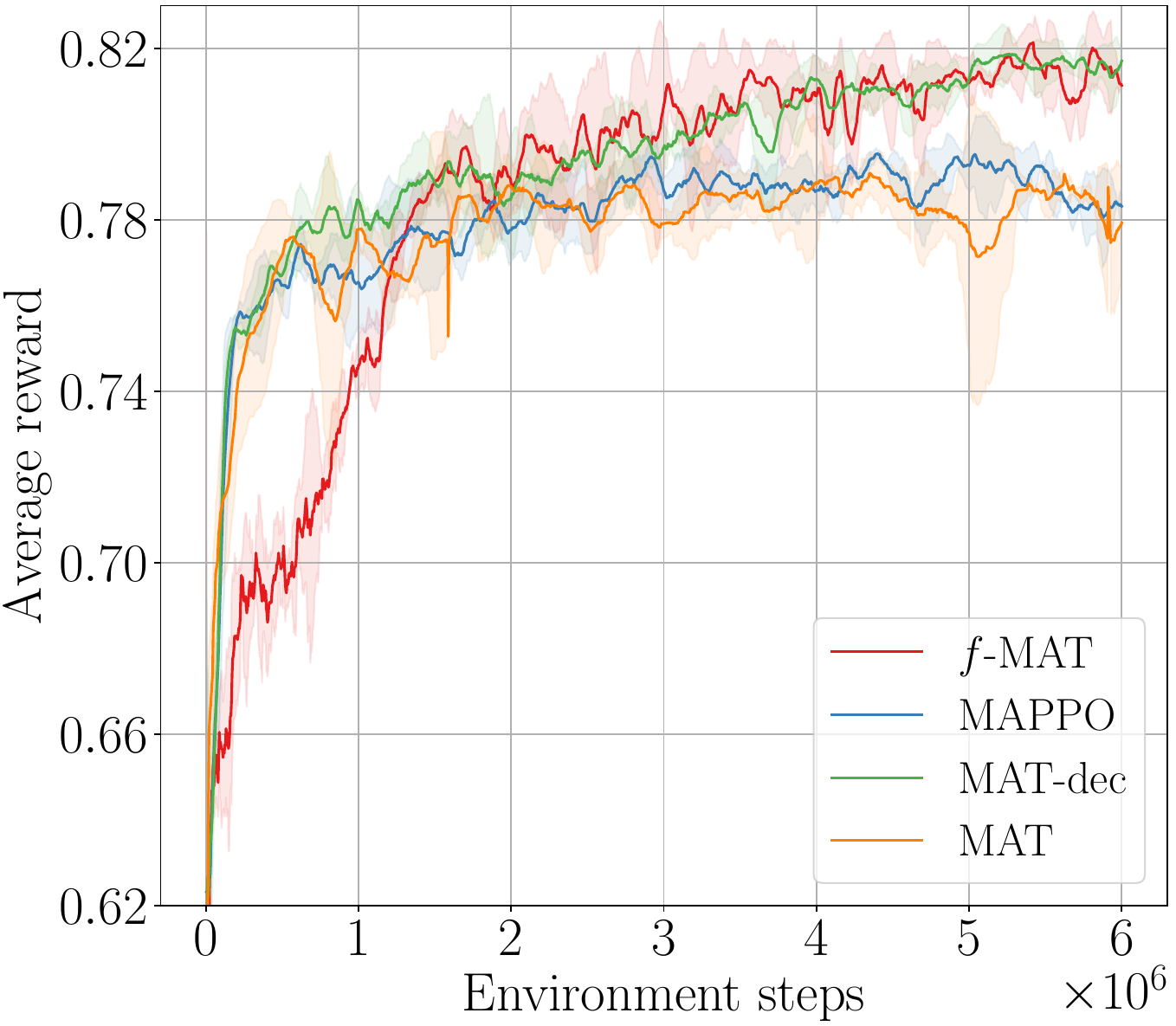} 
        \caption{Performance on microgrid-20}
        \label{fig:power20_per}
    \end{subfigure}%
    \hfill
    \begin{subfigure}[b]{0.23\textwidth}
        \centering
        \includegraphics[width=0.9\linewidth]{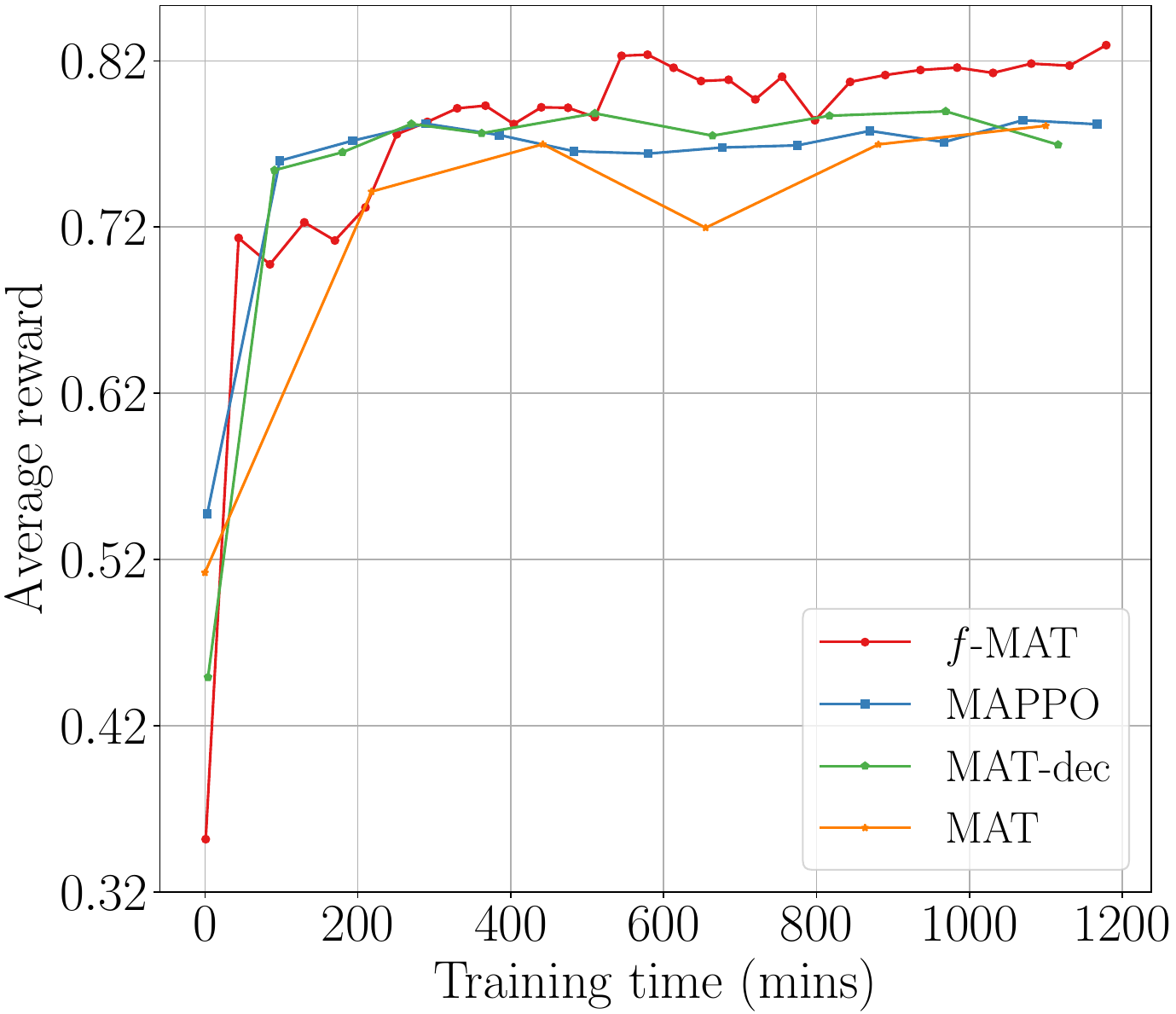} 
        \caption{Efficiency of microgrid-20}
        \label{fig:power20_efficiency}
    \end{subfigure}
    \caption{The performance and training efficiency results for \textbf{power grid control}.
    $f$-MAT outperforms more evidently in complex environments.}
    \label{fig:power_result}
\end{figure}


\noindent \textbf{Results} \quad  
We manually divided groups with only one overlapping agent.
In microgrid-6, we set the number of factors to $m=2$ and $S_f=4$.
In mircrogrid-20, we set the number of factors to $m=3$ and $S_f=6$.
This results in a complexity of \(O(S_fL)\).
Fig.~\ref{fig:power_result} shows that $f$-MAT is one of the best performing methods in the two microgrid systems,
achieving the highest training efficiency. 

One observation from this environment is that MAT outperforms other methods in microgrid 6, 
but suffers the poorest performance in microgrid 20, 
probably due to the local control nature of the setting, 
where an agent does not require information from all others to make decisions. 
Utilizing fully centralized observations during execution in the decentralized environment may introduce irrelevant information, potentially impairing performance.
This can also explain that MAT-dec yields results close to MAT and $f$-MAT in the microgrid 6 system,
and achieves one of the best performance in the microgrid 20.

Another observation is that $f$-MAT demonstrates its superior training efficiency more slowly in microgrid 6 compared to microgrid 20, suggesting that $f$-MAT exhibits its advantages more easily in complex environments.

In this section, we conducted experiments on three environments with different communication scopes: direct preceding neighborhoods (grid alignment), 
local neighborhoods (power control)
and broader areas (traffic light control). 
The selection of factors ranges from a clearly defined formulation (grid alignment) and limited scope (power control) to random choices on a general graph (traffic light control). 
In addition, the observation and action spaces include both homogeneous and heterogeneous settings. 

\begin{figure}[t]
    \centering
    \begin{subfigure}[b]{0.23\textwidth}
        \centering
        \includegraphics[width=0.9\linewidth]{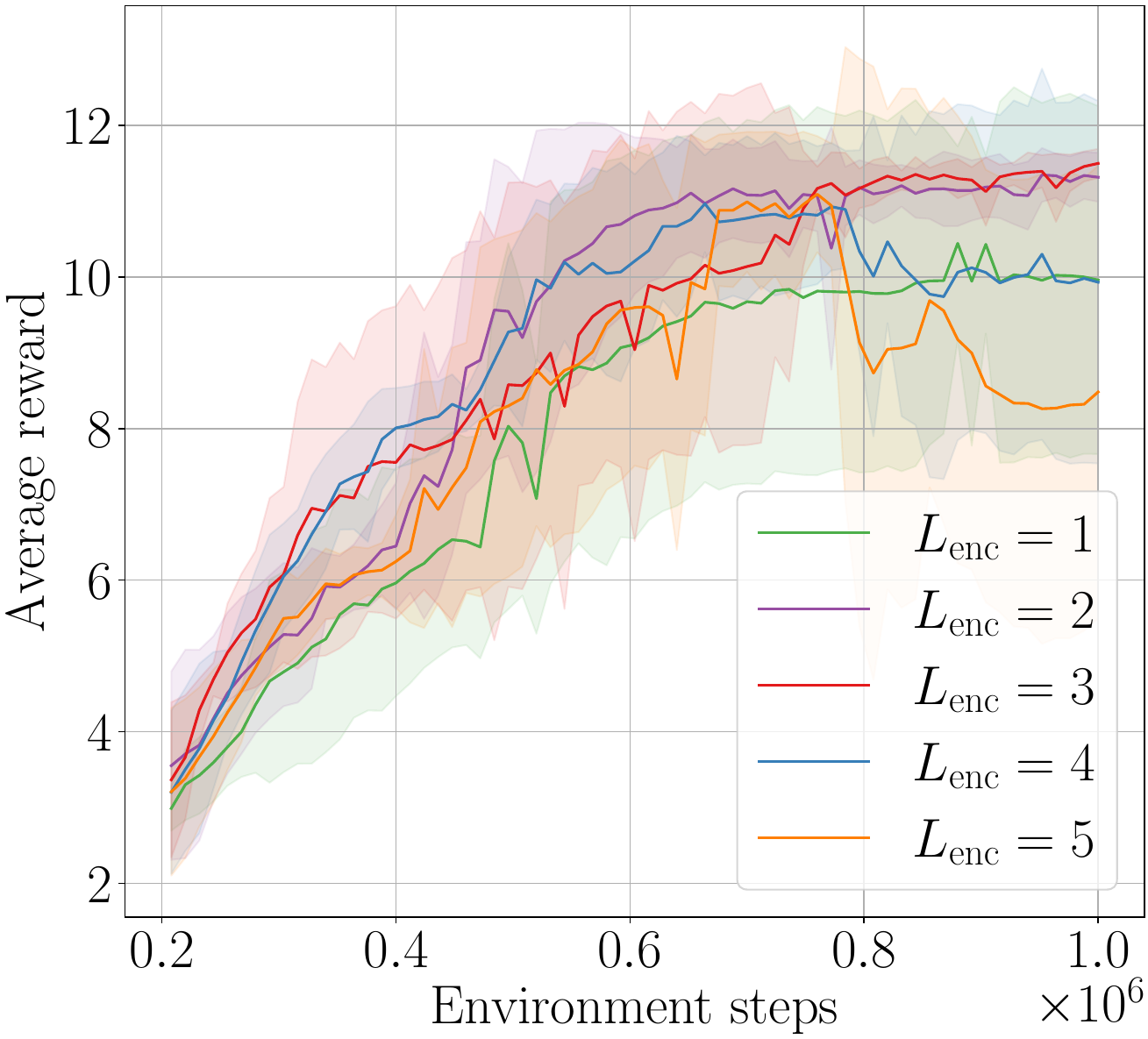} 
        \caption{Varying $L_\text{enc}$}
        \label{fig:abl_Lenc_gs6}
    \end{subfigure}%
    \hfill
    \begin{subfigure}[b]{0.23\textwidth}
        \centering
        \includegraphics[width=0.9\linewidth]{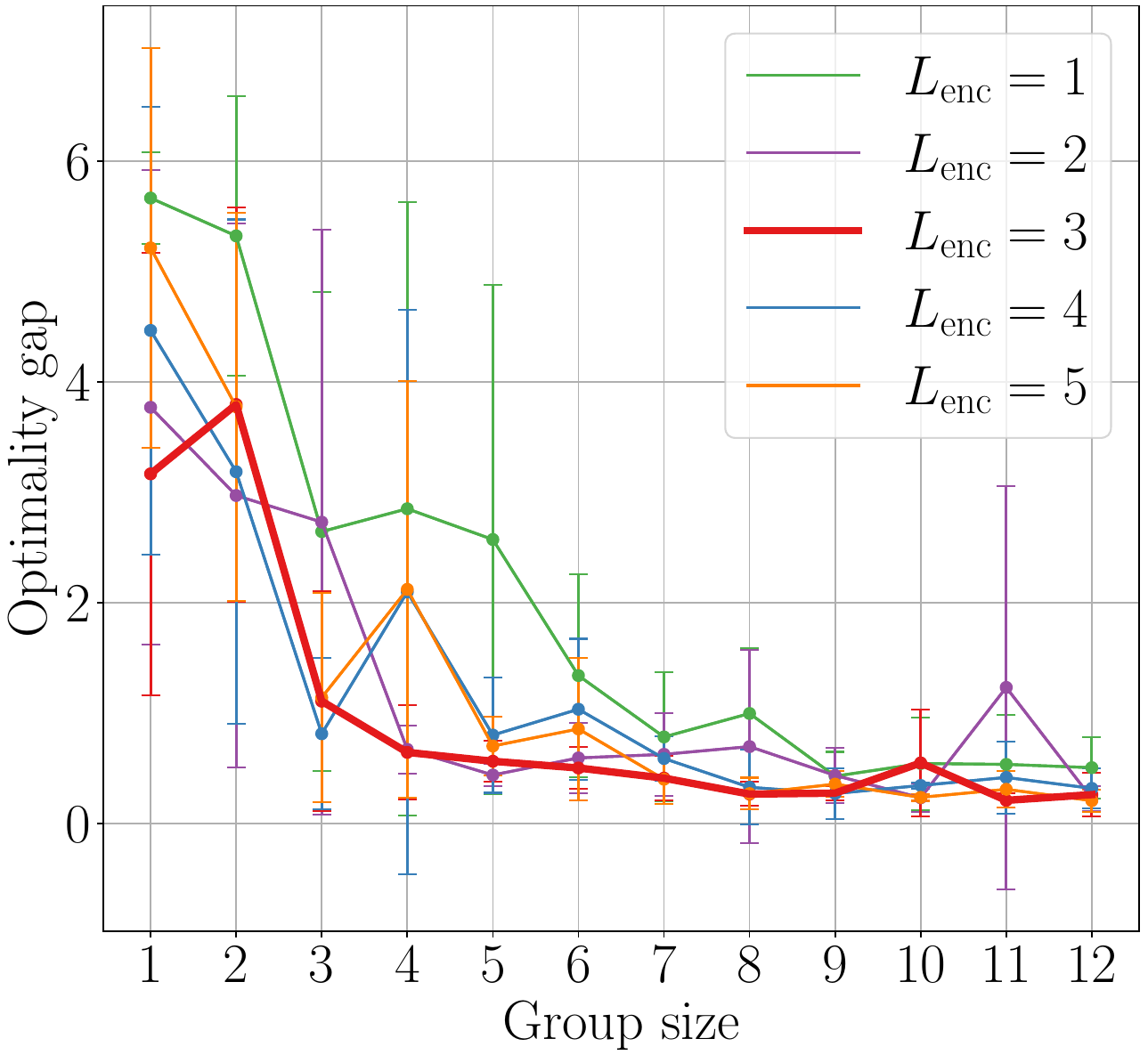} 
        \caption{Varying group size}
        \label{fig:abl_gs}
    \end{subfigure}
    \caption{Ablation on GridSim: (a) Average reward under different $L_\text{enc}$ with group size $=6$. (b) Comparison on optimality gap under different $L_\text{enc}$ and group sizes. }
    \label{fig:abl_grid}
\end{figure}

\subsection{Ablation}
\label{sec:alb}

We study the number of layers in $f$-MHA, group size and group selection in ablation as they are important components of our method.
Additionally, we substantiate our claim about the efficiency of $f$-MAT by comparing the computation time during inference with other baselines.

\noindent
\textbf{Choice of $L_{\text{enc}}$ and $L_{\text{dec}}$}  \quad  
We select a ${12 \times 12}$ grid from GridSim, fix the group size to $S_f = 6$, and then vary the value of $L_{\text{enc}}$ from 1 to 4. 
As shown in Fig. \ref{fig:abl_Lenc_gs6}, $L_{\text{enc}}=3$ produces the most stable trend and achieves the highest reward. The results of group sizes $S_f =9, 12$ in Appendix Fig. \ref{fig:supp_grid12_9} and \ref{fig:supp_grid12_12} further support this observation.
To explore the relationship between $L_\text{enc}$ and group size,
we use the optimality gap, the value between the true optimal reward and the learned reward achieved by the algorithm, to illustrate the variations.
In Fig. \ref{fig:abl_gs}, $L_\text{enc}=3$ generally yields good performance among various group sizes. 
We notice an initial improvement in performance as $L_\text{enc}$ increases. However, further increases do not consistently enhance results. 
As the number of layers grows, 
the information received by agents becomes more homogeneous, making it challenging to distinguish between individual agent features.
Such an over-smoothing effect is also observed in GNNs \citep{kipf2017semisupervised}. 

Furthermore, the results for group size $=12$ demonstrate that selecting the appropriate group can lead to optimal rewards with a smaller $L_\text{enc}$.
The results for group sizes from 6 to 11 show that slightly increasing $L_\text{enc}$ can also offset a less ideal group choice, but increasing it to 4 or 5 leads to limited benefits, as the occasional performance boost cannot compensate for the significantly increased computation time incurred by a larger value of $L_\text{enc}$. 

Based on the above experiments, we recommend setting \(L_\text{enc} = 3\), which we used to produce our main results. 

\noindent
\textbf{Choice of group size}  \quad  
It is challenging to theoretically analyze the impact of group selection and size.
Here, we quote some relevant results from Lemma 2 in \citet{qu2022scalable} and Theorem 3 in \citet{Ma2024} to support our results,
although their setting only considers direct links between agents rather than through factors.
As \citet{Ma2024} shows, the optimality gap decays exponentially with increasing $k$,
where $k$ is the number of hops used in learning.
But when $k$ is larger than a threshold, 
other aspects such as sample efficiency, computational cost, 
and the representational capacity of the neural network must be considered, leading to the increase of the optimality gap.

In our setting, the group size $S_f$ can be considered a counterpart of $k$.
So, the above theoretical results (for $k$) are consistent with our experiments with varied values of $S_f$ across different $L_\text{enc}$ shown in Fig. \ref{fig:abl_gs}.

To select $S_f$, an economical way is to choose the "elbow" value. 
Or, one can pick $S_f$ that achieves the best performance, provided that computational cost and efficiency are manageable.
As Fig. \ref{fig:abl_gs} shows,
with $L_\text{enc}=3$ in GridSim, the elbow value of 8 is an option, and the value of 11 performs the best.

\noindent
\textbf{Choice of group selection}  \quad  
The group selection in $f$-MAT is flexible.
In GridSim, we designated groups based on the definition of reward.
We can also first choose the number of groups
and then randomly pick agents and its related neighbor (directed neighbor, $k$-nearest neighbor, $k$-hop neighbor), similar to our approach with traffic lights.
From Fig.~\ref{fig:supp_monaco_op}, we selected 2 and 4 as number of factors discussed in Sec. ~\ref{sec:experiment_traffic}.

\noindent
\textbf{Computational time}  \quad  
Recall the complexity of MHA in $f$-MAT is $O(m \cdot S_f \cdot L)$.
We conducted experiments on the inference time cost to demonstrate this big-O complexity,
and the result in $12 \times 12$ GridSim is shown in the Table \ref{table:computation_time} in Appendix~\ref{sec:other_results_app}. 
The group size of $f$-MAT is 12.

$f$-MAT runs in a comparable computation time to MAPPO during inference, benefiting from its parallel action generation mechanism.
MAT generates actions autoregressively during inference, making it the slowest method during inference. 
MAT-dec modifies the decoder of MAT by replacing the attention block with an MLP, yet it continues to generate actions autoregressively during inference. 
As a result, MAT-dec is only slightly faster than MAT. 

\noindent
\textbf{Conclusion} \quad
In this paper, we propose Factor-based Multi-Agent-Transformer that enables efficient collaborations in both training and execution through all agents via graph modeling within the CTDE framework. 
This approach enriches CTDE framework by incorporating neighborhood interactions during execution.
Empirical results demonstrate that $f$-MAT achieves strong performance across diverse environments. 
Future work will concentrate on dynamic graphs and the development of learnable factors.

\bibliography{aaai25}

\appendix
\onecolumn
\section{Detailed pseudo-algorithms}
\label{sec:pseudo}
\begin{algorithm*}[ht]
    \caption{The entire $f$-MAT algorithm}
    \label{alg:ours_overall}
    \begin{algorithmic}[1]
    \Require Number of agents $n$, number of factors $f$, steps per episode $T$, number of minibatch $M$, number of rollouts $R$, number of time steps $S$. Episodes $K=S/(TR)$, minibatch size $B=RT/M$, number of PPO epoches $P$.
    \For{$k = 0, \ldots, K-1$}
      \For{$r = 0, \ldots, R-1$ (in parallel)}
        \For{$t = 0, 1, \ldots, T-1$}

        \State Collect a sequence of observation $o_{t}^{i_1}, \ldots, o_{t}^{i_n}$ from environments.

        \LineComment{\textcolor{blue}{\textbf{Inference Phase}}}
        \State  \multiline{
        Generate observation representation sequence $\hat{o}_{t}^{i_1}, \ldots, \hat{o}_{t}^{i_n}, \dots, \hat{o}_{t}^{i_{n+f}}$ by feeding observations to the encoder. The input of encoder should be $o_{t}^{i_1}, \ldots, o_{t}^{i_n}, \ldots, o_{t}^{i_{n+f}}.$}

        \State \multiline{
        Input $\hat{o}_{t}^{i_1}, \ldots, \hat{o}_{t}^{i_n}, \dots, \hat{o}_{t}^{i_{n+f}}$ to the decoder, then generate the actions $a_{t}^{i_1}, \ldots, a_{t}^{i_n}$ in parallel.}
        
        \State Execute joint action $a_{t}^{i_1}, \ldots, a_{t}^{i_n}$ in environments and collect the reward $R(\mathbf{o}_t, \mathbf{a}_t)$.
        \State \multiline{
        Insert $ \left( \mathbf{o}_t, \mathbf{a}_t, R(\mathbf{o}_t, \mathbf{a}_t) \right)$ in to replay buffer $\mathcal{B}$. $\mathbf{o}_t=(o_{t}^{i_1}, \ldots, o_{t}^{i_n})$, which is the raw observation. $\mathbf{a}_t = (a_{t}^{i_1}, \ldots, a_{t}^{i_n}$), which is the generated action.}
        \EndFor
      \EndFor
        \State Compute value function prediction $V_{\bar{\phi}}(\hat{\mathbf{o}}_{t+1})$.
        \State Compute the joint advantage function $\hat{A_t}$ via GAE.
        \State Compute return to go $R(\mathbf{o}_t, \mathbf{a}_t) = \hat{A}_t  + V_{\bar{\phi}}(\hat{\mathbf{o}}_{t+1})$.
        \LineComment{\textcolor{blue}{\textbf{Training Phase}}}
        \For { $\_$ in $P$ epochs}
            
            \State Sample a random minibatch of $B$ steps from $\mathcal{B}$.
            \For{each sample in the minibatch $B$}
            \State \multiline{
                Extend $o^{i_1}, \ldots, o^{i_n}$ to $o^{i_1}, \ldots, o^{i_n}, \ldots, o^{i_{n+f}}$ by averaging the observation of the related agents and put them into the encoder to get $\hat{o}^{i_1}, \ldots, \hat{o}^{i_n}, \ldots, \hat{o}^{i_{n+f}}$.}
                \State Generate $V_{\phi}(\hat{o}^{i_1}), \ldots, V_{\phi}(\hat{o}^{i_n})$ with the output layer of the encoder.
                \State Calculate $L_{\text{Encoder}(\phi)}$ with Equation \ref{eq:fmat_encoder}.

                \State \multiline{
                Input $\hat{o}^{i_1}, \ldots, \hat{o}^{i_n}, \ldots, \hat{o}^{i_{n+f}}$ to the decoder, and generate $\pi_{\theta}^{i_1}, \ldots, \pi_{\theta}^{i_n}$ in parallel. 
                } 
                \State Calculate $L_{\text{Decoder}(\theta)}$ with Equation \ref{eq:decoder} based on $\pi_{\theta}^{i_1}, \ldots, \pi_{\theta}^{i_n}$.
                \State \multiline{
                Update the encoder and the decoder by minimising $L_{\text{Encoder}(\phi)}+L_{\text{Decoder}(\theta)}$ with gradient descent.}
            \EndFor
        \EndFor
    \EndFor

\end{algorithmic}
\end{algorithm*}
\begin{algorithm*}[!ht]
    \caption{Encoder to compute the observation embeddings of all agents using self-attention only}
    \label{alg:encoder}
\begin{algorithmic}[1]
    \State Initialize $O[\mathcal{N},:]$ to raw observation representations.
For $j \in \mathcal{F}$, set $O[j,:]$ to the average value of $\{O[i,:] : i \in f_j \}$.
    
    \For{$l = 1, 2, \ldots, L_{\text{enc}}$}
        
        
        \State $O \gets O + \textbf{MHA}(O, M_{n \to f})$  

        \State
        $O[t,:] \gets $ layer\_norm($O[t,:]$) for all $t \in \mathcal{N}  \cup \mathcal{F}$ 

        \State $O \gets O + \textbf{MHA}(O, M_{f \to n})$ 

        \State $O[t,:] \gets $ layer\_norm($O[t,:]$) for all $t \in \mathcal{N}  \cup \mathcal{F}$

        \State $O[\mathcal{N} \cup \mathcal{F},:] \gets O[\mathcal{N} \cup \mathcal{F},:] + \textbf{MLP}(O[\mathcal{N} \cup \mathcal{F},:])$

        \State $O[t,:] \gets $ layer\_norm($O[t,:]$) for all $t \in \mathcal{N}  \cup \mathcal{F}$
        
    \EndFor
    \Ensure $\hat{o}^t \gets \hat{O}[t,:]$ \ \ for all $t \in \mathcal{N}  \cup \mathcal{F}$.
\end{algorithmic}
\end{algorithm*}
\clearpage
\begin{algorithm*}[!ht]
    \caption{Decoder to compute the action of all agents in parallel}
    \label{alg:decoder}
\begin{algorithmic}[1]
    \Require $\hat{O}[\mathcal{N} \cup \mathcal{F}, :]$ which is the \textbf{observation} representation from the result of the encoder.
    \For{$i = 1, 2, \ldots, n$ (in parallel)} 
    \LineComment{\textcolor{red}{\textbf{Parallel action initialization}}}
    \State Initialize action representation $A[\mathcal{N} \cup \mathcal{F}, :]$ by MLP($\hat{O}[\mathcal{N} \cup \mathcal{F}, :]$). 
    
    \For{$l = 1, 2, \ldots, L_{\text{dec}}$}
        
        
        \State $A \gets A + \textbf{MHA}(A, M_{n \to f})$

        \State
        $A[i, :] \gets $ layer\_norm($A[t, :]$) for all $i \in \mathcal{N}  \cup \mathcal{F}$ 

        \State $A \gets A + \textbf{MHA}(A, M_{f \to n})$

        \State $A[i, :] \gets $ layer\_norm($A[i, :]$) for all $t \in \mathcal{N}  \cup \mathcal{F}$

        \State $A \gets \hat{O} + \textbf{MHA}(\hat{O}, A, M_{n \to f})$

        \State $A[i, :] \gets $ layer\_norm($A[i, :]$) for all $t \in \mathcal{N}  \cup \mathcal{F}$

        \State $A \gets \hat{O} + \textbf{MHA}(\hat{O}, A, M_{f \to n})$

        \State $A[i, :] \gets $ layer\_norm($A[i, :]$) for all $t \in \mathcal{N}  \cup \mathcal{F}$

        \State $A[\mathcal{N}  \cup \mathcal{F}, :] \gets A[\mathcal{N}  \cup \mathcal{F}, :] + \textbf{MLP}(A[\mathcal{N}  \cup \mathcal{F}, :])$

        \State $A[i, :] \gets $ layer\_norm($A[i, :]$) for all $t \in \mathcal{N}  \cup \mathcal{F}$
    \EndFor

        \LineComment{\textcolor{red}{\textbf{Parallel action generation}}}
        \State Sample $a^i$ from a categorical distribution based on MLP($\hat{a}^i$) 
    \EndFor

    \Ensure $\hat{a}^i \gets A[i, :]$ (embedding vector) and $a^i$ (scalar action)\ \ for all $i \in \mathcal{N}$
\end{algorithmic}
\end{algorithm*}
\begin{algorithm*}
    \caption{Parallel Action Generation}
    \label{alg:parallel_sampling}
\begin{algorithmic}[1]
    \Require $\hat{O}[\mathcal{N} \cup \mathcal{F}, :]$ the \textbf{observation} representation from the output of the encoder.
    \For{$i = 1, 2, \ldots, n$ (in parallel)} 
        \LineComment{\textcolor{red}{\textbf{Parallel action initialization}}}
        \State {
        Initialize action representation 
        $A[\mathcal{N} \cup \mathcal{F}, :]$ by MLP($\hat{O}[\mathcal{N} \cup \mathcal{F}, :]$)}
        
        \For{$l = 1, 2, \ldots, L_{\text{dec}}$}
            \State $A \gets A + \textbf{MHA}(\hat{A}, M_{n \to f})$ 
            \State $A \gets A + \textbf{MHA}(A, M_{f \to n})$
            \State $A \gets \hat{O} + \textbf{MHA}(\hat{O}, A, M_{n \to f})$ 
            \State $A \gets \hat{O} + \textbf{MHA}(\hat{O}, A, M_{f \to n})$
        \EndFor
        \LineComment{\textcolor{red}{\textbf{Parallel action generation}}}
        \State Sample $a^i$ from a categorical distribution based on MLP($\hat{a}^i$) 
    \EndFor

    \Ensure $\hat{a}^i \gets A[i, :]$ (embedding vector) and $a^i$ (scalar action)\ \ for all $i \in \mathcal{N}$
    
\end{algorithmic}
\end{algorithm*}

\section{Inspiration from Gibbs sampling}
\label{sec:gibbs_app}

We revisit the idea of Gibbs sampling~\citep{casella1992explaining} in graphical model, where one would like to sample from a set of factored probability distributions (as it is typically difficult to sample from the joint distribution). Although directly sampling from the joint distribution is hard, one could show that Gibbs sampling coverage to the right stationary distributions under certain conditions~\citep{geman1984stochastic}. This setting aligns with our setting where we would like to sample actions from factored graph of agents while directly sampling is difficult. We could therefore stack multiple attention layers to practically implement multiple iterations of Gibbs sampling steps. Note that Gibbs sampler can be easily parallelized (simultaneous sampling of all variables), although at the cost of being not ergodic~\citep{newman2007distributed, gonzalez2011parallel}.

\section{Environment Illustration}
\label{sec:app_env}

\begin{figure*}[htbp] 
    \centering
    \begin{subfigure}[b]{0.2\textwidth}
        \centering
        \includegraphics[width=0.9\textwidth]{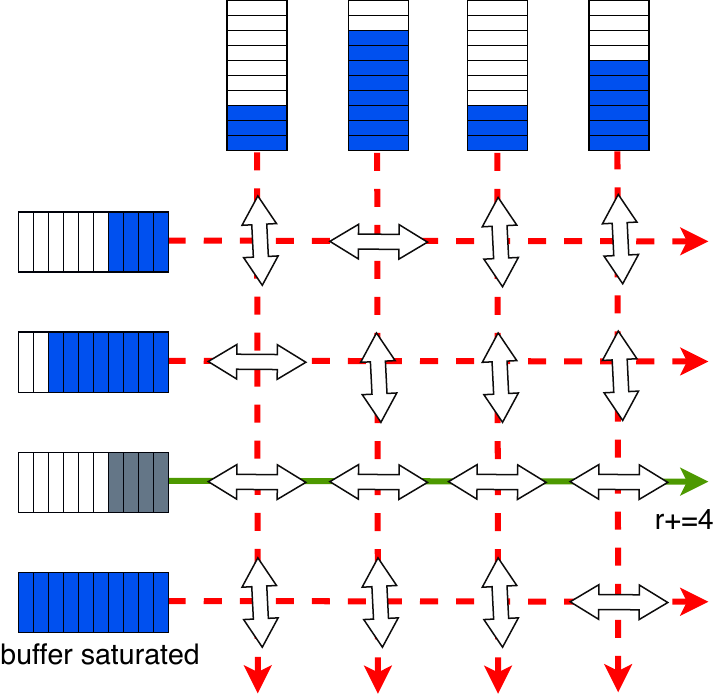}
        \caption{Grid Alignment}
        \label{fig:env_gridsim}
    \end{subfigure}
    \hfill 
    \begin{subfigure}[b]{0.15\textwidth}
        \centering
        \includegraphics[width=1.3\textwidth]{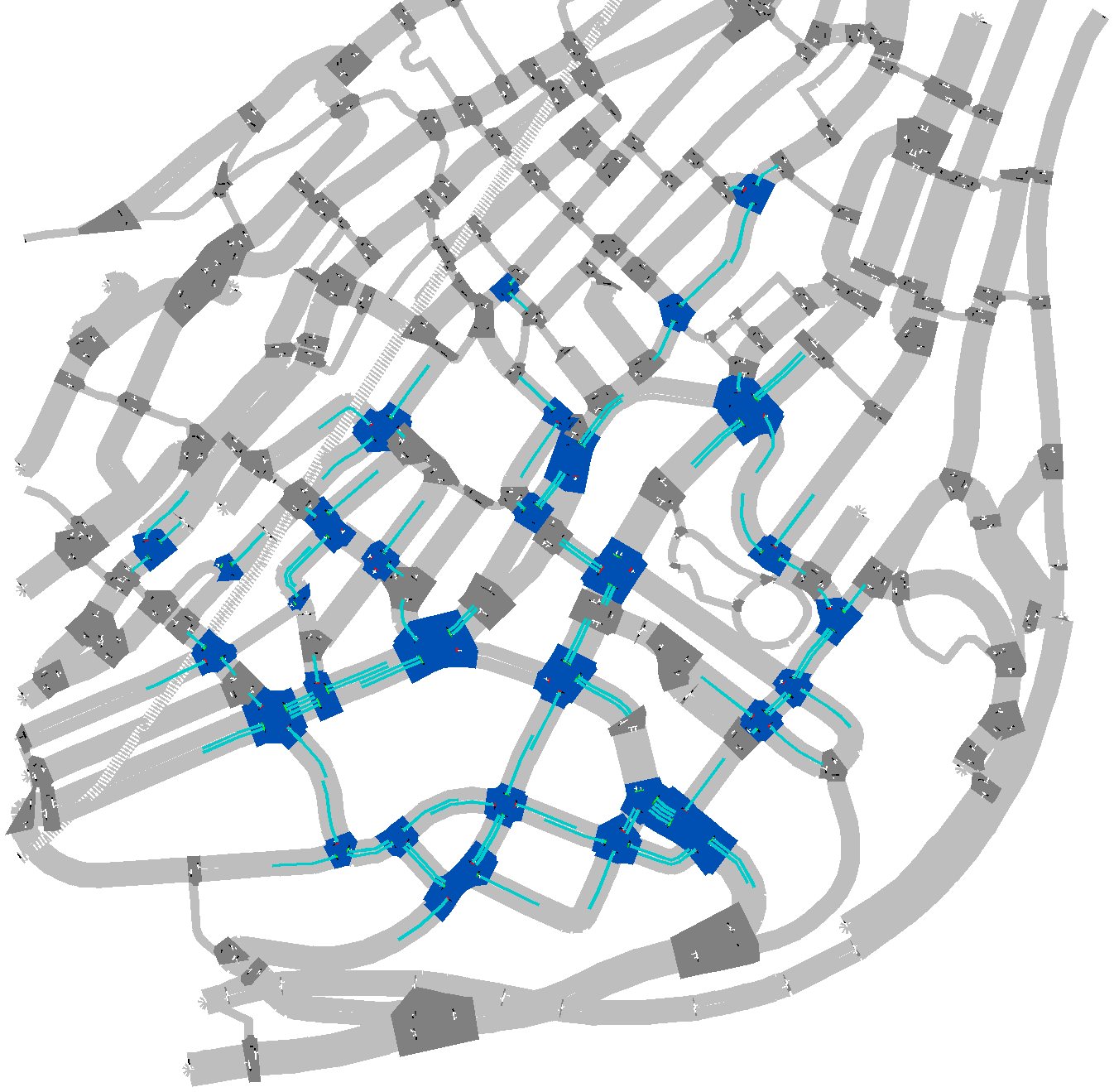}
        \caption{Monaco}
        \label{fig:env_monaco}
    \end{subfigure}
    \hfill 
    \begin{subfigure}[b]{0.5\textwidth}
        \centering
        \includegraphics[width=\textwidth]{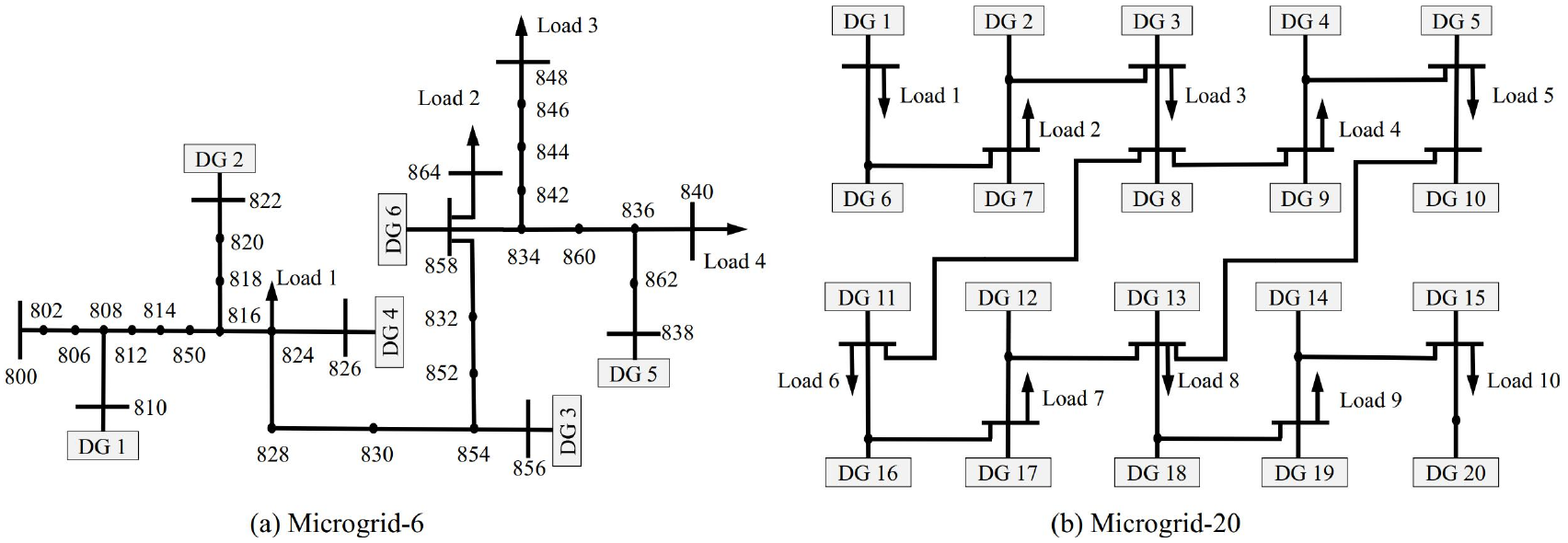}
        \caption{PowerGrid}
        \label{fig:env_power}
    \end{subfigure}
    \caption{Illustration of three environments. (a) is borrowed from
    \citet{zhang2007conditional}.
    (b) is borrowed from \citet{chu2020multi}.
    (c) is borrowed from \citet{chen2021powernet}.}
    \label{sec:env_illu}
\end{figure*}

\paragraph{Grid Alignment}
The details on grid alignment are provided in Sec. ~\ref{sec:grid_align}. 
To elaborate on this environment, we run Fig.\ref{fig:env_gridsim} as an example.
Along the row with the grey buffer, all gates are oriented horizontally, allowing all four traffic units in the buffer to pass through this row, thereby increasing the global reward by 4.

\paragraph{Monaco} 
The blue area in Fig. \ref{fig:env_monaco} represents a real-world traffic network consisting of 28 intersections from the Monaco city. 
The reward for each agent is calculated as the total of queue lengths across all incoming lanes.

\paragraph{PowerGrid} 
Figure \ref{fig:env_power} illustrates the structures of 6 distributed DGs (microgrid-6) 
and a larger-scale microgrid system with 20 DGs (microgrid-20).
To better simulate the real-world system, we introduce random disturbances at each simulation step, varying within $\pm5$\% of the nominal values for each load.

\section{Supplementary for Experiment}
\label{sec:other_results_app}

\subsection{Grid Alignment}
\begin{figure}[htbp] 
    \centering
    
    \begin{subfigure}[b]{0.45\textwidth} 
        \centering
        \includegraphics[width=0.45\linewidth]{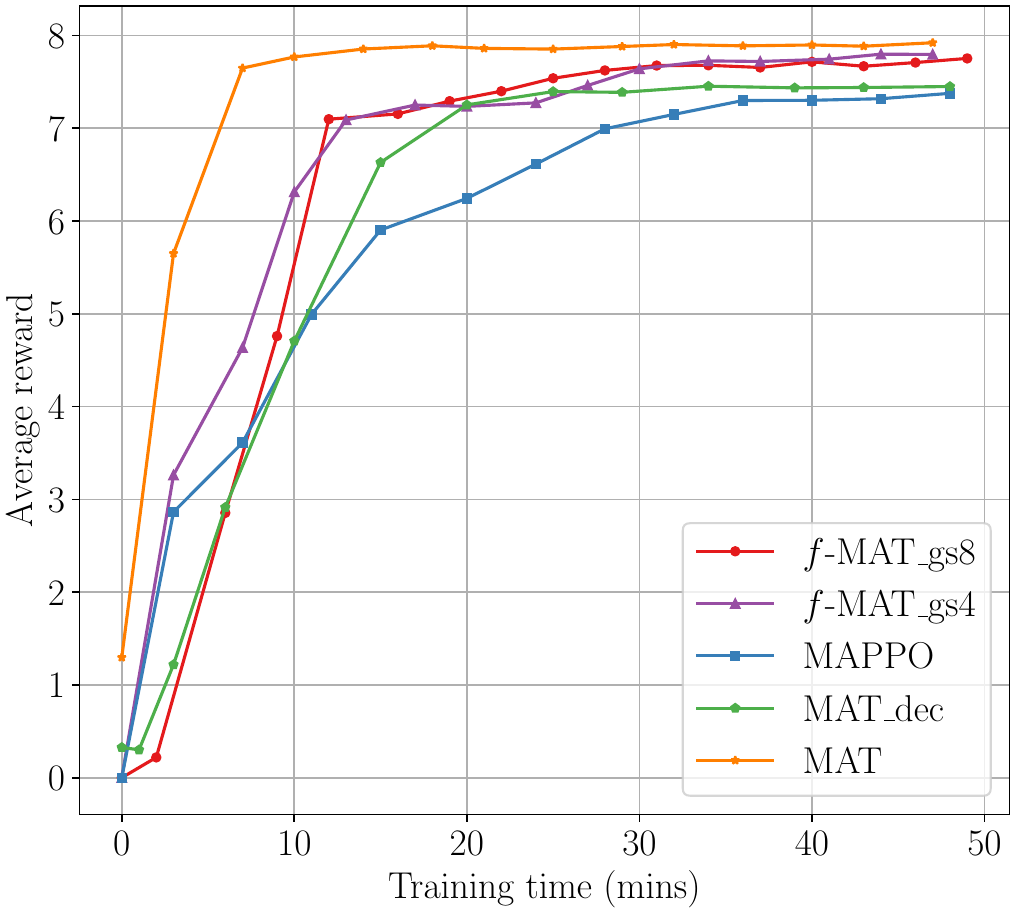}
        \quad 
        \includegraphics[width=0.45\linewidth]{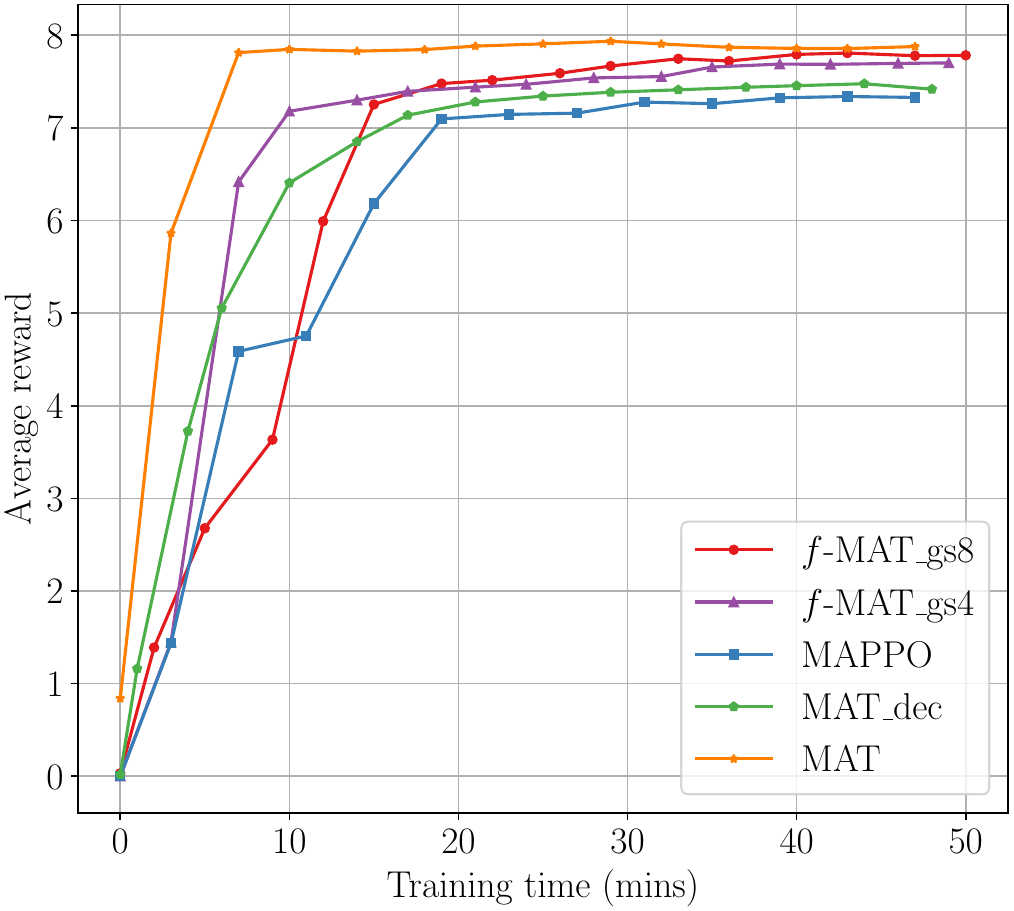}
        \caption{Efficiency in gridsim $8 \times 8$}
        \label{fig:supp_grid8_time}
    \end{subfigure}%
    \hfill 
    \begin{subfigure}[b]{0.24\textwidth}
        \centering
        \includegraphics[width=0.9\linewidth]{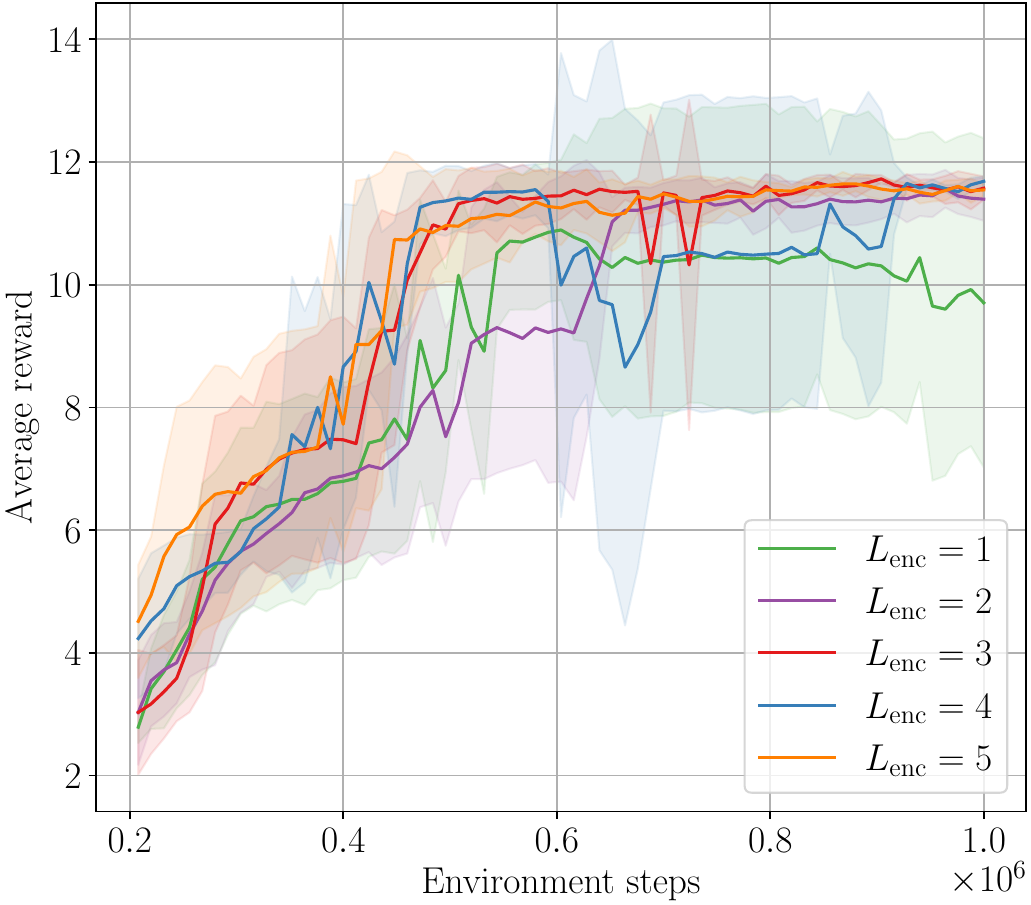}
        \caption{Group size $=9$}
        \label{fig:supp_grid12_9}
    \end{subfigure}
        \hfill 
    \begin{subfigure}[b]{0.24\textwidth}
        \centering
        \includegraphics[width=\linewidth]{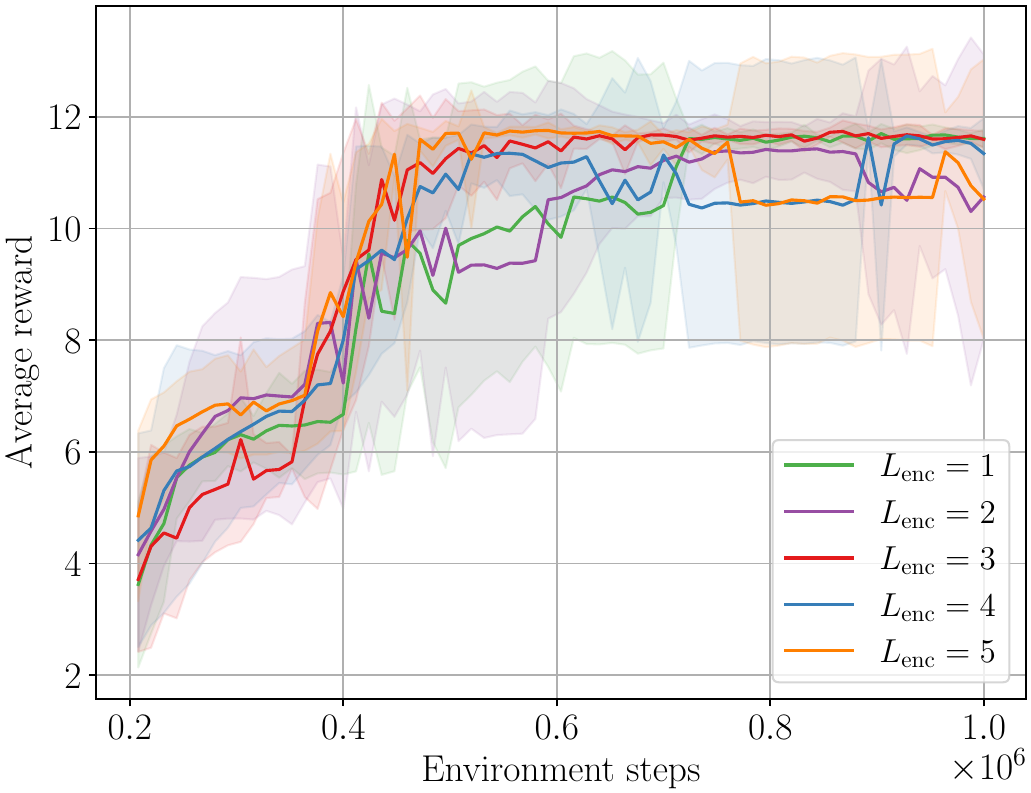}
        \caption{Group size $=12$}
        \label{fig:supp_grid12_12}
    \end{subfigure}
    \caption{Supplementary for GridSim: (a) Efficiency in gridsim $8 \times 8$ for two more seeds. (b) and (c) varied $L_\text{enc}$ across different group sizes, proving that $L_\text{enc}=3$ is the appropriate choice.} 
    \label{fig:supp_time_grid_traffic}
\end{figure}

\subsection{Traffic Light Control}
\begin{figure}[!htbp] 
    \centering
    \begin{subfigure}[b]{0.6\textwidth}
        \centering
        \includegraphics[width=0.45\linewidth]{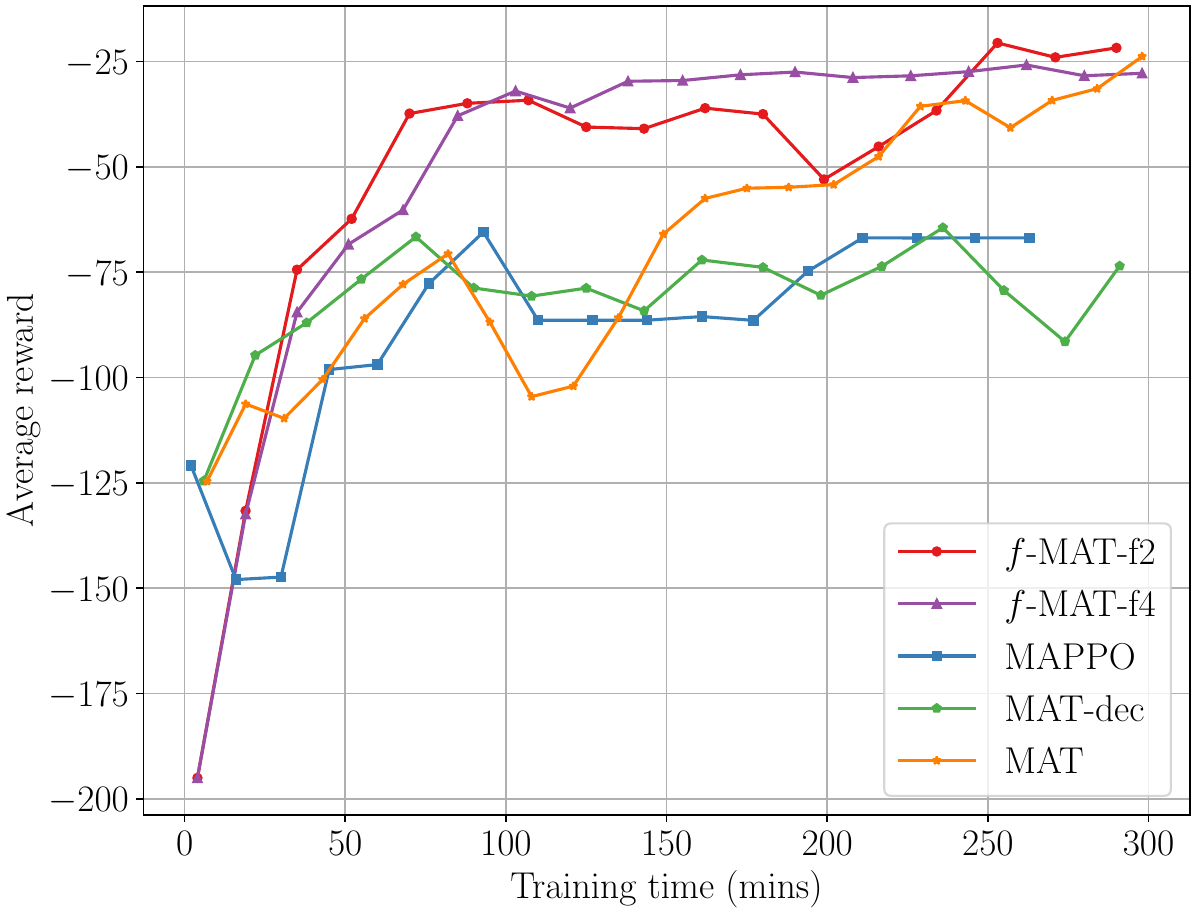}
        \quad 
        \includegraphics[width=0.45\linewidth]{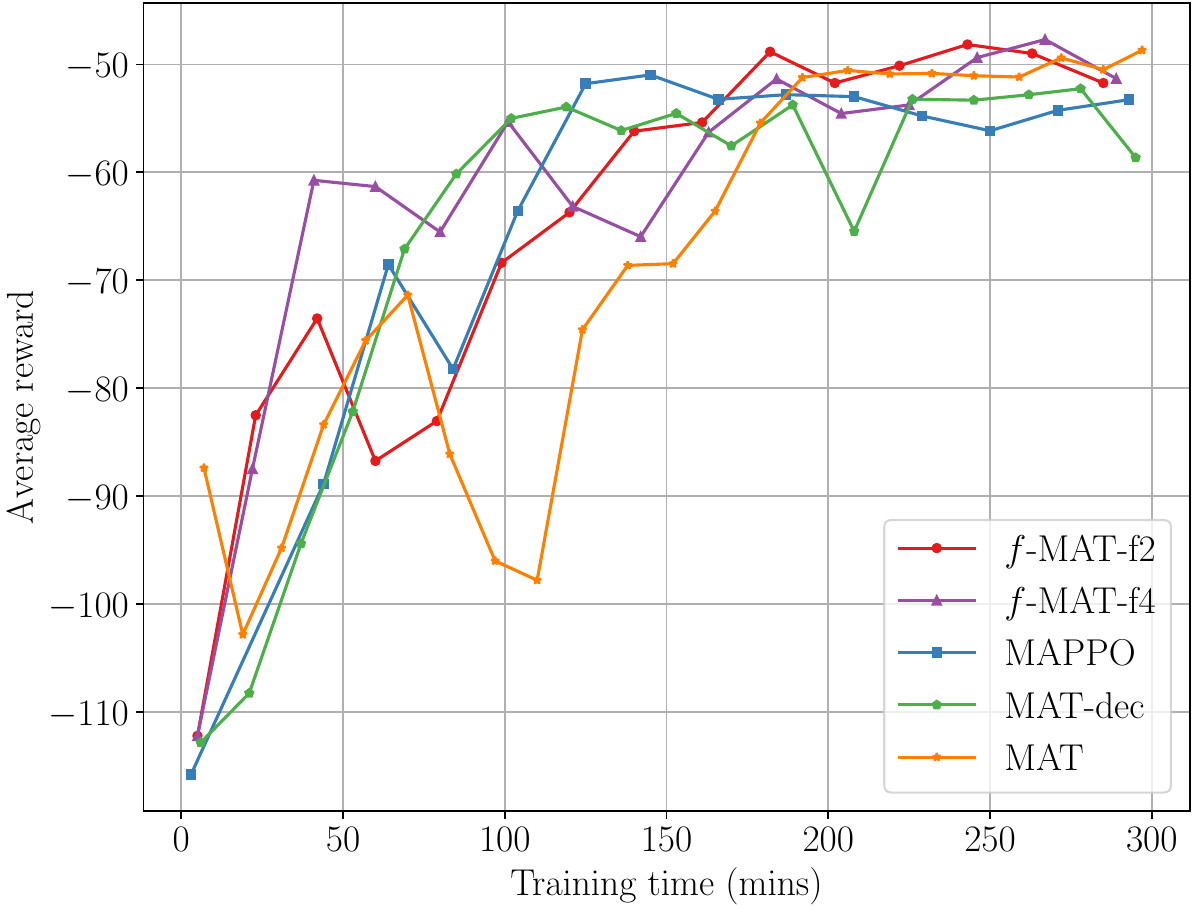}
        \caption{Efficiency in Monaco}
        \label{fig:supp_monaco_time}
    \end{subfigure}
    \hfill
    \begin{subfigure}[b]{0.3\textwidth} 
        \centering
        \includegraphics[width=0.9\linewidth]{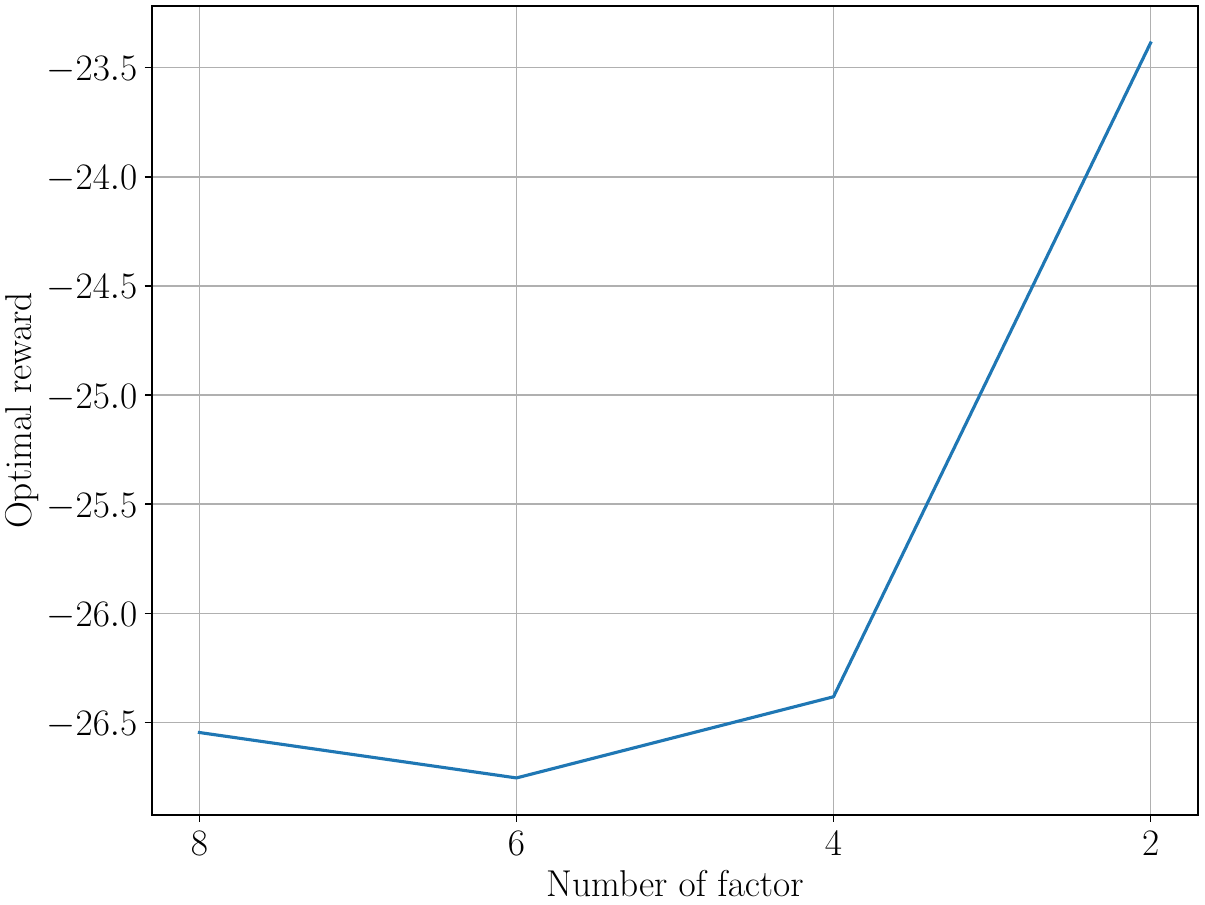}
        \caption{Varying number of factors}
        \label{fig:supp_monaco_op}
    \end{subfigure}%
    \caption{Supplementary for training efficiency and optimality gap in Monaco. 
    Fig.~\ref{fig:supp_monaco_time} shows the efficiency under two more seeds.
    In Fig \ref{fig:supp_monaco_op}, we illustrated the optimal reward achieved under different number of factors with $L_\text{enc}=3$, showing an exponential growth as the number of factor reduces. We selected the turning point with the number of factors equals to 2, and the optimal point with the number of factors equals to 4, as the results discussed in Sec. ~\ref{sec:experiment_traffic}.} 
    \label{fig:supp_time_power}
\end{figure}
\clearpage

\subsection{Power Control}
\begin{figure}[!ht] 
    \centering
    \begin{subfigure}[b]{0.5\textwidth} 
        \centering
        \includegraphics[width=0.45\linewidth]{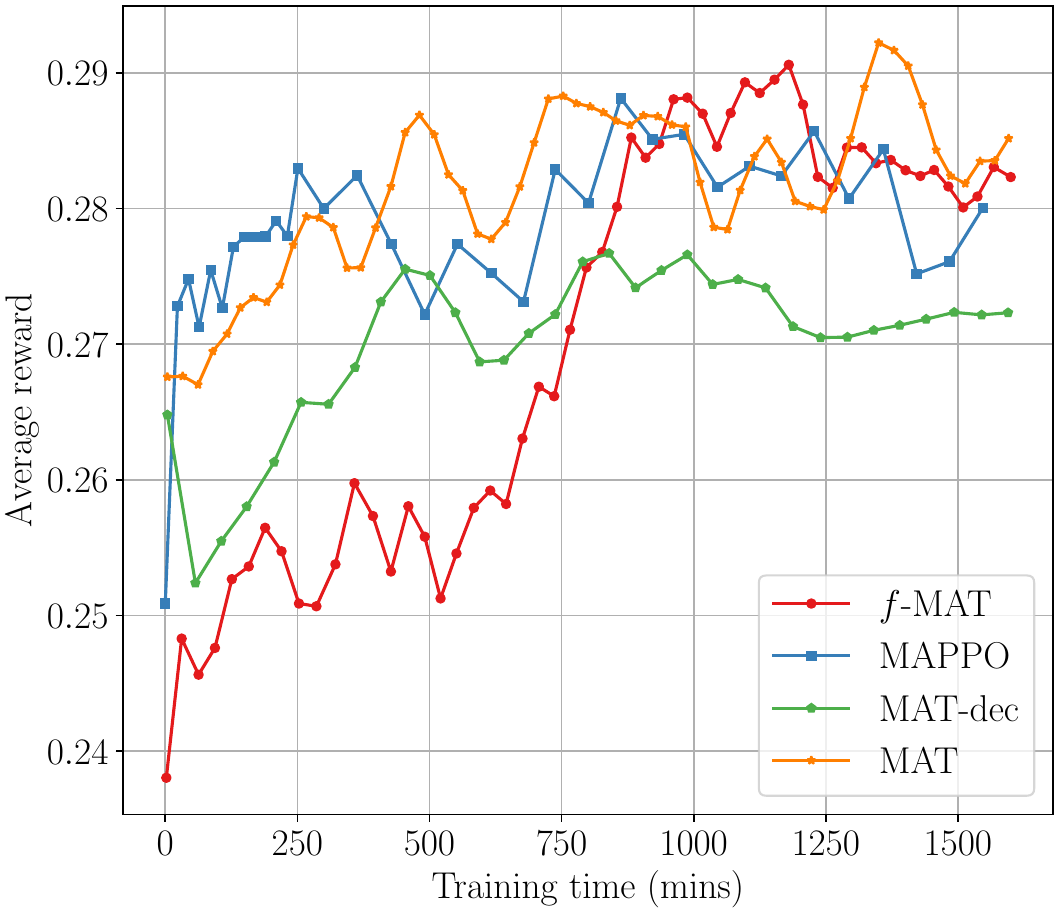}
        \quad 
        \includegraphics[width=0.45\linewidth]{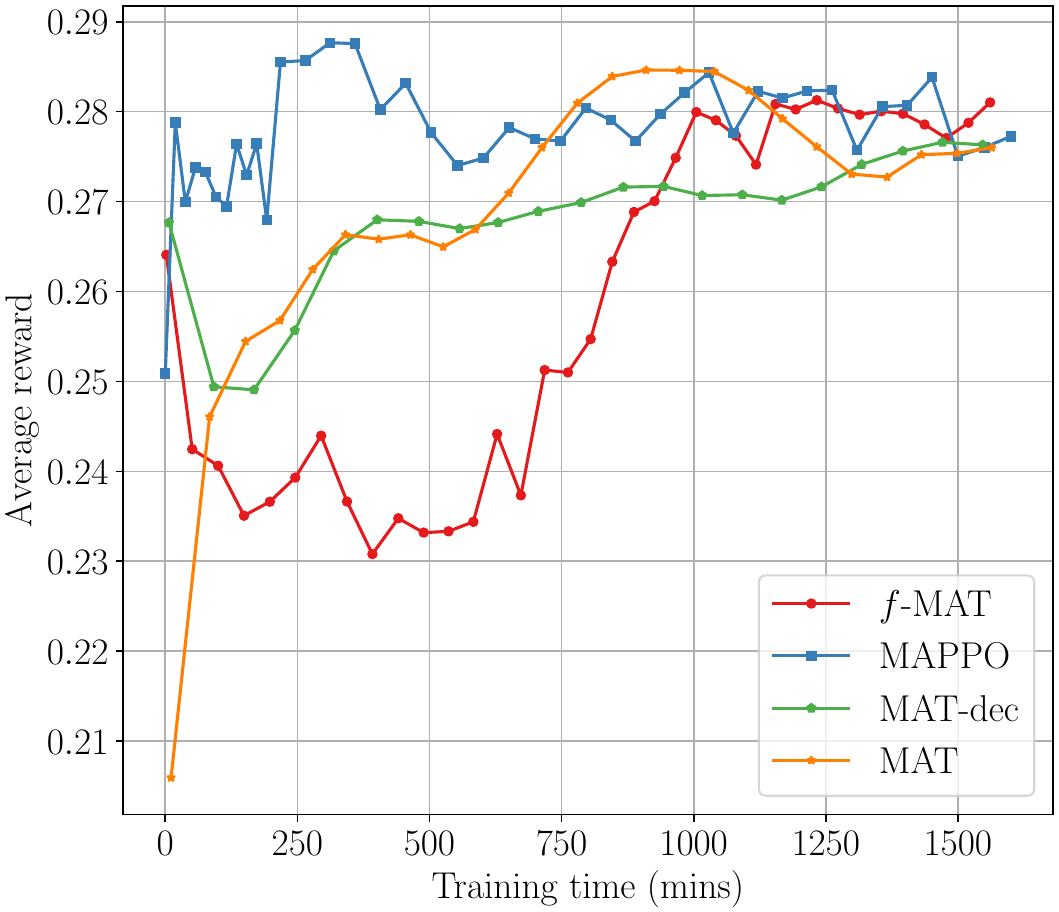}
        \caption{Efficiency in microgrid-6.}
        \label{fig:supp_power6_time}
    \end{subfigure}%
    \hfill 
    \begin{subfigure}[b]{0.5\textwidth}
        \centering
        \includegraphics[width=0.45\linewidth]{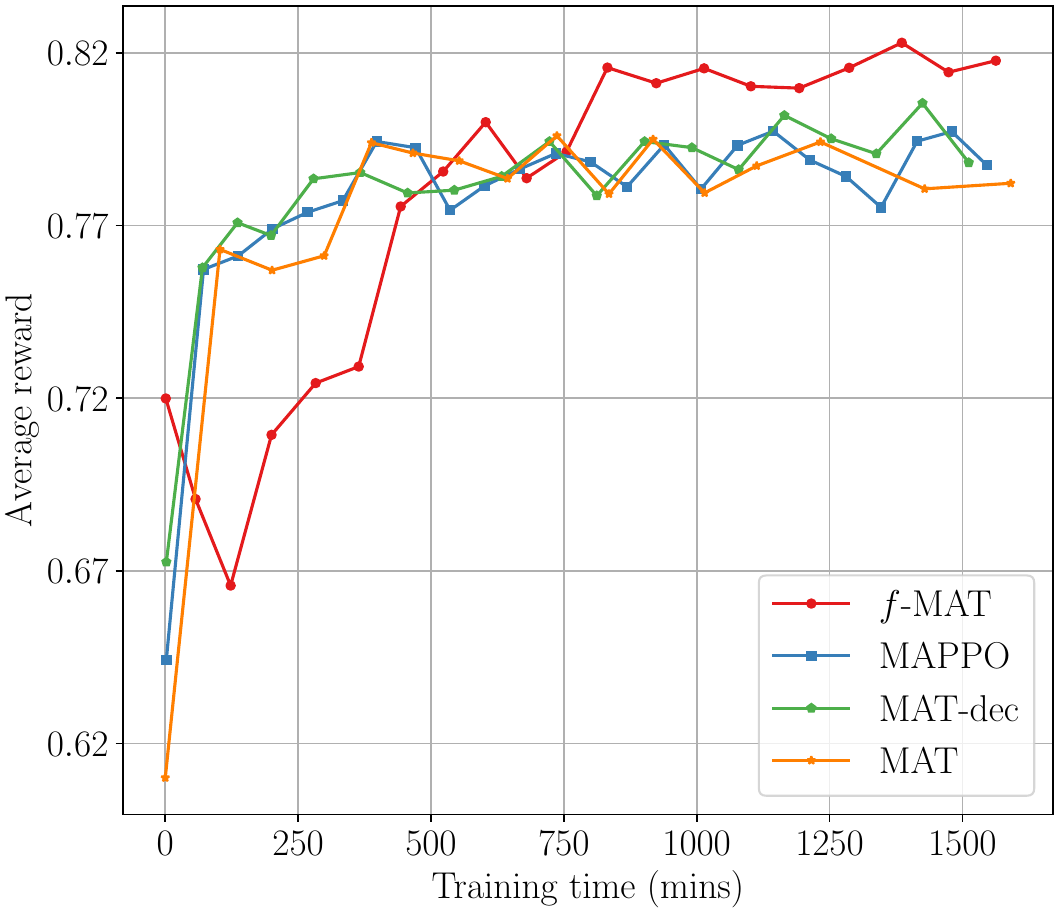}
        \quad 
        \includegraphics[width=0.45\linewidth]{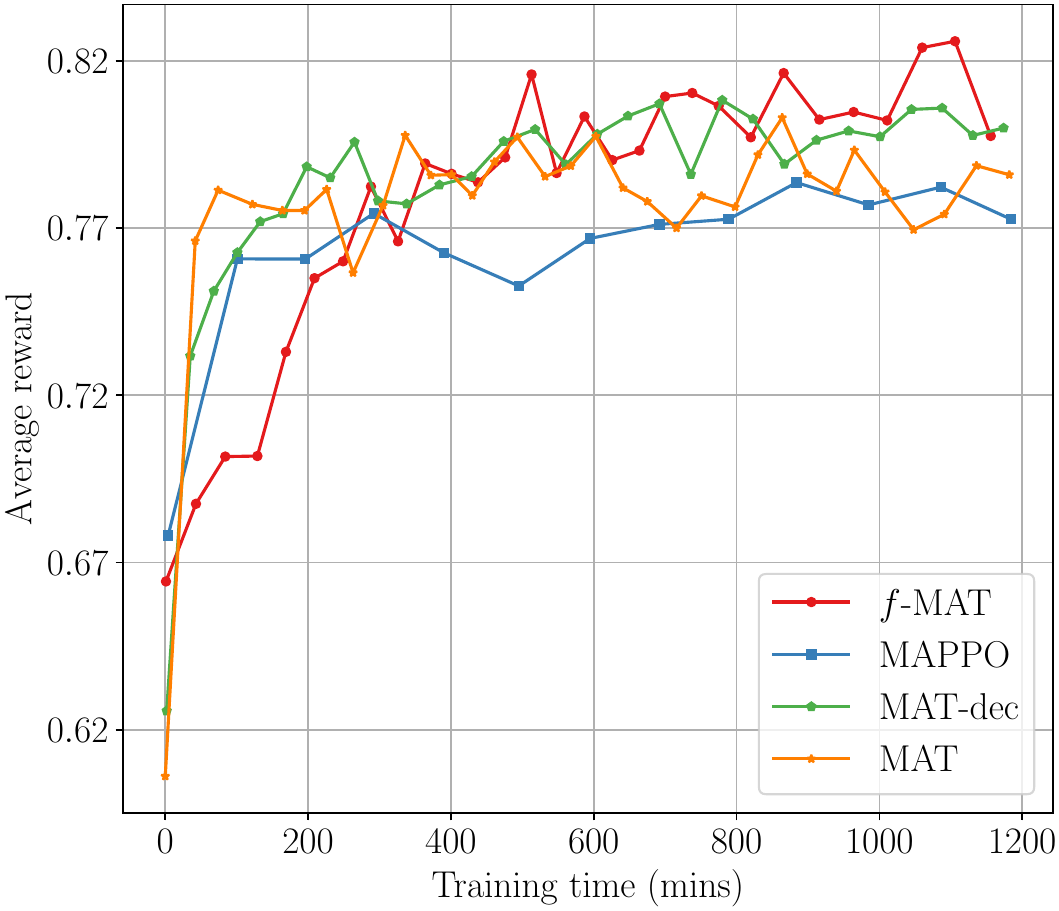}
        \caption{Efficiency in microgrid-20.}
        \label{fig:supp_power20_time}
    \end{subfigure}
    \caption{Supplementary with two more seeds on training efficiency in power control illustrates the superiority of $f$-MAT in complex environments with a larger number of agents.} 
    \label{fig:supp_power_time}
\end{figure}
\begin{table}[!ht]
\centering
\caption{Computation time of different methods running on $12 \times 12$ GridSim. Group size of $f$-MAT is 12.}
\label{table:computation_time}
\begin{tabular}{c!{\vrule width \lightrulewidth}cl} 
\cmidrule[\heavyrulewidth]{1-3}
Method    & Inference (s) &   \\ 
\cmidrule{1-3}
$f$-MAT-$L_{\text{enc}=1}$            & $0.0096$            &   \\ 
\cmidrule{1-3}
$f$-MAT-$L_{\text{enc}=3}$          & $0.01488$          &   \\ 
\cmidrule{1-3}
MAT            & $0.2695$          &   \\ 
\cmidrule{1-3}
MAT-dec               & $0.1639$          &   \\
\cmidrule{1-3}
MAPPO               & $0.0051$          &   \\
\cmidrule[\heavyrulewidth]{1-3}
\end{tabular}
\end{table}

\begin{table}[!ht]
\centering
\caption{The performance comparison between $f$-MAT and communication-based methods draws on results from \cite{chen2021powernet}. This includes the fully centralized method CommNet, the fully decentralized method ConseNet, and the CTDE method DIAL.
The results show that $f$-MAT significantly outperforms these communication-based methods.}
\label{table:power}
\begin{tabular}{c!{\vrule width \lightrulewidth}c!{\vrule width \lightrulewidth}cl} 
\cmidrule[\heavyrulewidth]{1-3}
Network   & Mircogrid-6 & Mircogrid-20 &   \\ 
\cmidrule{1-3}
$f$-MAT & $0.291_{\pm 0.089\%}$           & $0.8315_{\pm 0.2\%}$            &   \\ 
\cmidrule{1-3}
ConseNet  & $0.221_{\pm 0.16\%}$         & $0.681_{\pm 2.58\%}$          &   \\ 
\cmidrule{1-3}
CommNet   & $0.221_{\pm 0.14\%}$          & $0.680_{\pm 2.21\%}$          &   \\ 
\cmidrule{1-3}
DIAL      & $0.222_{\pm 0.04\%}$         & $0.689_{\pm 1.85\%}$          &   \\
\cmidrule[\heavyrulewidth]{1-3}
\end{tabular}
\end{table}


\end{document}